\documentclass[twocolumn,showkeys,preprintnumbers,amsmath,amssymb]{revtex4}

\usepackage{bm}
\usepackage{graphicx}

\usepackage{epstopdf}

\usepackage{psfrag}  
\usepackage{amsmath,amsthm}
\usepackage{bm}

\begin{document}

\preprint{APS}

\title{On the origins of Hierarchy in Complex Networks}

\author{Bernat Corominas-Murtra$^{1,2,4}$ , Joaqu\'in Go\~ni$^{3}$, Ricard V. Sol\'e$^{2,4,5}$, Carlos Rodr\'iguez-Caso$^{2,4}$
}
\vspace{0.25 cm}
\affiliation{
$^1$ Section for Science of Complex Systems; Medical University of Vienna, Spitalgasse 23; A-1090, Austria\\
$^2$ ICREA-Complex Systems Lab, Universitat Pompeu Fabra, 08003 Barcelona, Spain\\
$^3$ Dpt. of Psychological and Brain Sciences. Indiana University. Bloomington, IN 47405\\
$^4$ Institut de Biologia Evolutiva, CSIC-UPF, Passeig Mar\'itim de la Barceloneta, 37, 08003 Barcelona, Spain\\
$^5$ Santa Fe Institute, 1399 Hyde Park road, Santa Fe NM 87501, USA}

\begin{abstract}
 Hierarchy seems to pervade complexity in both living and artificial systems. Despite its relevance, no general theory that captures all features of 
hierarchy and its origins has been proposed yet. 
Here we present a formal approach resulting from the convergence of theoretical morphology and network theory that allows constructing a 3D 
morphospace of hierarchies and hence comparing the hierarchical 
organization of ecological, cellular, technological and social networks. Embedded within large voids in the morphospace of all possible hierarchies, 
four major groups are identified. Two of them match the expected 
from random networks with similar connectivity, thus suggesting that non-adaptive factors are at work. 
Ecological and gene networks define the other two, indicating that their topological 
order is the result of functional constraints. These results are consistent with an 
exploration of the morphospace using in silico evolved networks.
\end{abstract}


\maketitle
\section{introduction}
Fifty years ago \cite{Simon:1962}, Herbert Simon defined complex systems as nested 
hierarchical networks of components organized as interconnected modules.  Hierarchy seems a pervasive 
feature of the organization of natural and artificial systems \cite{Mihm2010,Stanley:1996}. The examples span from
social interactions \cite{Guimera:2003, Valverde2007}, urban growth \cite{Krugman1996, Batty1994}, and allometric scaling \cite{West1997} to cell 
function \cite{Ma2004a,Yu2006,Lagomarsino2007,Nitin:2010,Rodriguez-Caso:2009}, development \cite{Erwin2009}, ecosystem flows \cite
{Hirata1985,Wickens1988}, river networks \cite{iturbe1997}, brain organization \cite{Kaiser2010} and macroevolution \cite{Eldredge1985, 
McShea2001}. But hierarchy is a polysemous word, involving order, levels, inclusion or control as possible 
descriptors \cite{Lane2006}, none of which captures either its complexity or the problem of its measure and origins. 
Although previous work using complex networks theory has quantitatively tackled the problem \cite{Guimera:2003, Ravasz2002, Vazquez2002,
Trusina2004, Clauset2008a, Corominas-Murtra:2011,Dehmer2008, rammal:86, Song2006, Nicolis1986, Vicsek2012}, some questions remain: 
Is  hierarchy a widespread feature of complex systems organization? What types of hierarchies do exist?  Are hierarchies 
the result of selection pressures or, conversely, do they arise as a byproduct of structural constraints?

A well established concept where such 
questions are addressed involves the use of a {\em morphospace} \cite{Niklas:1994, McGhee1999, Thomas:2000, Shoval2012, Schuetz2012} namely a 
phenotype space where a small set of quantitative traits can be defined as the axes. Here we take a step in this direction by combining 
morphospace and network theories, taking the intuitive idea of hierarchy as the starting point: A pattern of relations where there is no ambiguity in who 
controls whom with a pyramidal structure in which the few control the many. Formally,  the picture of hierarchy matches a tree of relations \cite
{Whyte1969}, ideally represented by a directed graph. As shown in Fig. (\ref{Fig:fig_1}) the elements of the system are represented by nodes connected 
by arrows establishing the map of relations of who affects whom. Accordingly, a measure of hierarchy should account for the deviations from this ideal 
tree picture. Such deviations occur because:  (a) several elements are on the top, (b) downstream elements interact horizontally or (c) feedback loops 
are present. As shown in Fig. (\ref{Fig:fig_1}a), the tree-like picture matches the concept of genealogies, taxonomies, armies and corporations. Conversely, drainage 
networks in river basins \cite{iturbe1997} would define a reverse anti-hierarchical situation, as depicted in Fig. (\ref{Fig:fig_1}b). In this structure, multiple 
elements on the top merge downstream like an inverted tree. In between them, we can place a more or less symmetric web (Fig. \ref{Fig:fig_1}c) 
somewhat combining both tendencies. But, in general, neither biological nor technological webs match the feedforward pattern. In most real systems, signal 
integration requires gathering inputs from different  sources, while robust processing and control requires crosstalk and feedbacks, as represented by 
Fig. (\ref{Fig:fig_1}d), which are often organized in a modular fashion \cite{Kitano2004a}. 

Hierarchy seems to pervade a coherent form of organization that allows reducing the costs associated to reliable 
information transmission \cite{Guimera:2001} and to support efficient genetic and metabolic control in cellular networks \cite{Stelling2004}. However, 
a unified picture of hierarchy should not only provide a formal definition but also help understanding 
the forces that shape it. Here we provide the formalization and the quantitative characterization of the morphospace of the possible hierarchies. The study of a large number of real networks and its comparison with model systems provide some unexpected answers to the previous questions.

\section{The coordinates of Hierarchy}

\begin{figure*}
\begin{center}
	\includegraphics[width=13.5 cm]{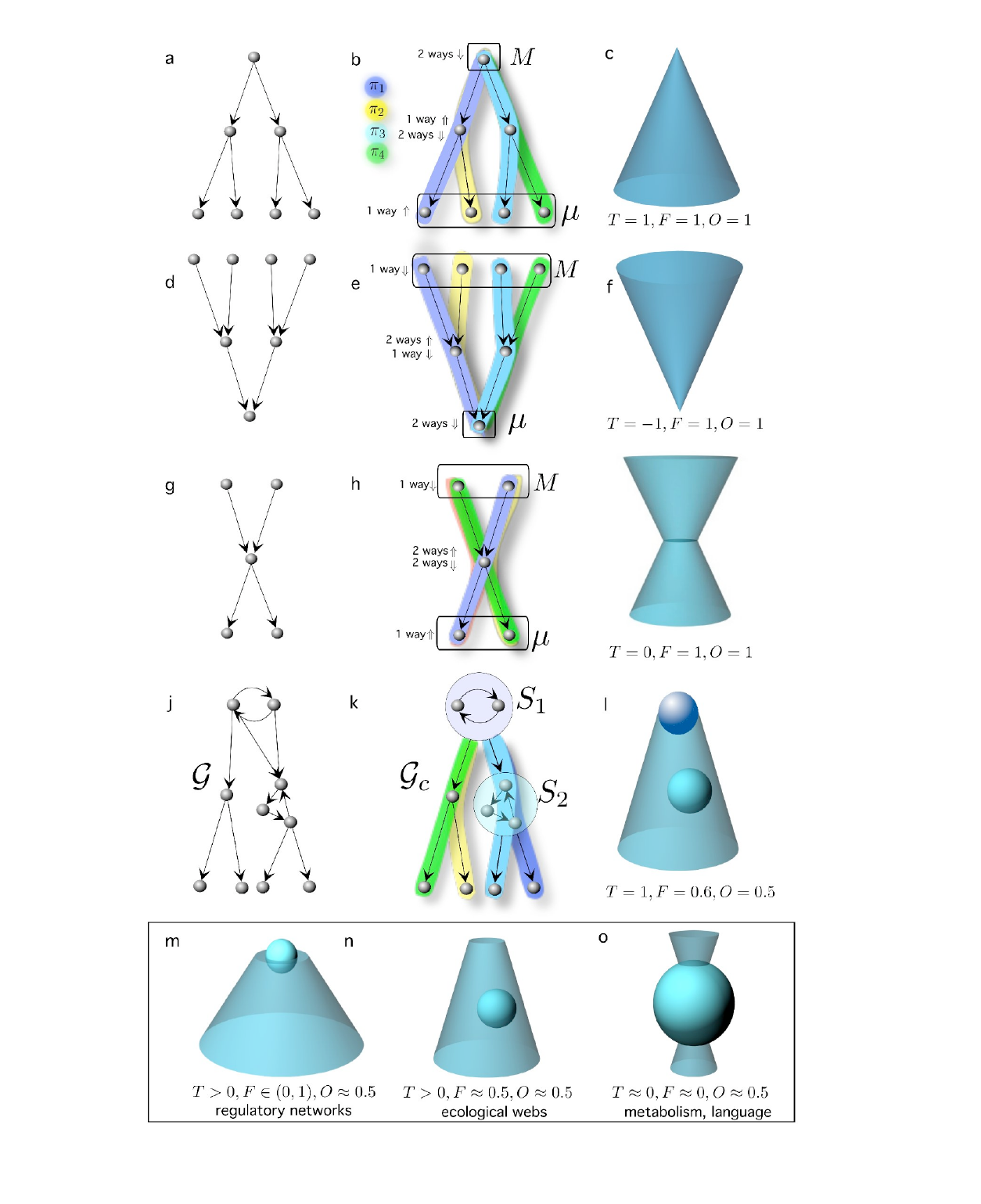}
	\caption{Defining network hierarchy. The graphical representation of a tree-like 
hierarchical graph (a), an inverted tree matching the anti-hierarchical graph (b), a non-hierarchical graph (c) with feedforward structures and a graph  $
{\cal G}$ displaying cycles (d). In (e-f), all of them present $4$ pathways $(\pi_1,\pi_2,\pi_3,\pi_4)$ from maximals -$M$ (top nodes)- to minimals -$\mu$ 
(bottom nodes)-. 
In  (e), downstream diversity of paths is $H_f({\cal G})=\log 4$, contrasting with the absence of uncertainty when reversing them $H_b({\cal G})=0$. In (f) 
the behavior is exactly the opposite: $H_f({\cal G})=0$ and $H_b({\cal G})=\log 4$. In graph (g), the non-hierarchical feedforward structure presents the 
same forward and backward uncertainties, $H_f({\cal G})=H({\cal G})=\log 2$.
Chart (h) represents the node weighted condensed graph, ${\cal G}_{\cal C}$, computed by first detecting the $SCC$s and then collapsing every $S_i$ 
into single nodes. For $S_1$ and $S_2$, the node weights are $\alpha_1=2$ and $\alpha_2=3$, respectively. With a representative icon involving 
cones and balls (i-l) charts show $TFO$ show the values for (a-d) graphs. In  a vertical box, icons corresponding to three representative systems: gene 
regulatory networks (m), ecological networks (n) and metabolism and language (o). Icons are provided as an intuitive picture on how the flow is 
organized within the network. They are used in fig. 2 for a more comprehensive view of the morphospace behavior.}
\label{Fig:fig_1}
\end{center}
\end{figure*}

Since hierarchy is about relations, our  approach formalizes the interaction between system's elements by means of a directed graph ${\cal G}(V,E)$ 
\cite{Gross:1998} where $v_i \in V$ ($i=1, ..., n$) is a node and $\langle v_i,v_j\rangle \in E$ an arrow going from $v_i$ to $v_j$. This graph is 
transformed into the key object for our methodology: the so-called {\em node weighted condensed graph}, ${\cal G}_{\cal C}(V_{\cal C}, E_{\cal C})$. As 
Figs. (\ref{Fig:fig_1}d and \ref{Fig:fig_1}h) show,  ${\cal G}_{\cal C}$ is a feedforward structure where cyclic modules of ${\cal G}$ (the so-called strongly 
connected components $\{S_1, ... ,S_{\ell}\}$,  hereafter $SCC$s) are represented by individual nodes (thereby obtaining the {\em condensed graph} 
\cite{Gross:1998}). It is worth to mention that $SCC$ detection has been shown as a powerful approach for subsystem identification \cite{Bonchev:2005, 
Zhao:2006, Rodriguez-Caso:2009} unraveling the presence of nested organizations. In the case of a given ${\cal G}_{\cal C}$, every node has a weight  $\alpha_i$ that indicates the number of elements from ${\cal G}$ it represents (see section \ref{Sec:DAGs} in the Appendix for details). Interestingly, such a 
graph represents the largest set of subgraphs of ${\cal G}$ (which, in this case, turn to be either isolated nodes or $SCC$'s) that can be properly {\em 
ordered} attending the causal flow defined by the arrows. This latter property will be the key to conceptually connect our definition of hierarchy with 
order.

Our space of hierarchies will be a metric space $\Omega$, defined from three 
coordinates: treeness ($T$), feedforwardness ($F$) and orderability ($O$), which properly quantify graph hierarchy.

\subsection{Treeness}

Treeness ($T$, being $T$ in the range $-1 \le T \le 1$) weights how pyramidal is the structure and how unambiguous is its chain of command. This measure 
covers the range from hierarchical ($T>0$, Fig. (\ref{Fig:fig_1}a)) to anti-hierarchical ($T<0$, Fig. (\ref{Fig:fig_1}b)) graphs including those 
structures which do not exhibit any pyramidal behaviour ($T=0$, Fig. (\ref{Fig:fig_1}c)). As illustrated in Fig. (\ref{Fig:fig_1}e-g), these graphs are 
characterized by taking the structure as a road map where we compare the 
diversity of choices we can take going top-down, i.e., following the arrows of the structure, versus the uncertainty generated when reverting the paths going bottom-up. 
Such diversity is properly quantified using {\em forward} ($H_f$) and {\em backward} ($H_b$) entropies, respectively. Entropies are computed over 
directed acyclic graphs \cite{Corominas-Murtra:2011, Corominas-Murtra:2010}. While section \ref{Sec:DetailedDerivation} of the Appendix presents a rigorous description of these concepts, they 
can be briefly presented departing from the node-weighted condensed graph ${\cal G}_{\cal C}(V_{\cal C}, E_{\cal C})$ shown in Fig. (\ref{Fig:fig_1}h). Notice that a directed acyclic graph is naturally a node-weighted graph and hence ensures ${\cal G}={\cal G}_{\cal C}$. From the  graph ${\cal G}_{\cal C}$,
we define two sets, namely $M$ and $\mu$. The first, composed by nodes with $k_{in}=0$, i.e. the set of  {\em maximal} nodes and the second, the 
set of nodes with $k_{out}=0$, to be referred to as the set of {\em minimal nodes}. Now, let $\Pi_{M\mu}$ be the set of all paths starting in some 
maximal node. Since ${\cal G}_{\cal C}$ is a graph without cycles, this set contains a finite number of elements $\pi_1,. . ., \pi_N$. If $v_i\in M$, the 
uncertainty associated to follow a given path starting from $v_i$ and ending to some node in $\mu$, $h(v_i)$ will be:
\[
h(v_i)=-\sum_{\pi_k\in\Pi_{M\mu}}\mathbb{P}(\pi_k|v_i)\log \mathbb{P}(\pi_k|v_i),
\]
where $\mathbb{P}(\pi_k|v_i)$ is the probability that the path $\pi_k$ is followed, starting from node $v_i\in M$. The average uncertainty we face to 
follow a path starting from some node in $M$ will be 
$H_f({\cal G}_{\cal C})=\frac{1}{|M|}\sum_{v_i\in M}h(v_i)$
which is the general expression of the {\em forward} entropy of ${\cal G_C}$. 
In a similar way but reversing the pathways a {\em backwards} entropy, $H_b({\cal G}_{\cal C})$, of ${\cal G}_{\cal C}$ can be obtained. Details of the derivation of $H_b({\cal G}_{\cal C})$ are found in section \ref{Sec:DetailedDerivation} of  the Appendix.

Once the entropies accounting for the top-down and bottom-up path diversity generation are properly derived, we proceed to define a function with the aim of quantitatively grasp the deviations of ${\cal G}_{\cal C}$ from the ideal tree picture of a hierarchical system. 
The explicit form is obtained by means of the normalized difference \cite{Corominas-Murtra:2011} of the two presented entropies, namely:
\[
f({\cal G})=\frac{H_f({\cal G}_{\cal C})-H_b({\cal G}_{\cal C})}{\max\{H_f({\cal G}_{\cal C}),H_b({\cal G}_{\cal C})\}}.
\]
For the sake of consistency $f({\cal G})\equiv 0$ if $H_f({\cal G}_{\cal C})=H_b({\cal G}_{\cal C})=0$ -occurring in the case where ${\cal G}_{\cal C}$ is a linear chain.
The final value  $T({\cal G})$ is computed as follows: First, let ${\cal W}({\cal G})$  be set containing ${\cal G}_{\cal C}$ and all subgraphs of ${\cal G}_{\cal C}$ that can obtained by means of the application of a {\em leaf removal algorithm} (either top-down or bottom up, see section \ref{Sec:DetailedDerivation} of the Appendix for details). Then, $T({\cal G})$ is obtained by averaging $f$ along all the members of ${\cal W}({\cal G})$:
\begin{equation}
T({\cal G})=\langle f\rangle_{{\cal W}({\cal G})}.
\label{eq:T}
\end{equation}
In addition, $T({\cal G})\equiv 0$ if ${\cal G}_{\cal C}$ has no links -which can happen, e.g.,  if ${\cal G}$ is totally cyclic. $T({\cal G})$ has been shown to be a very good indicator of the deviations of a given acyclic graph from the ideal hierarchical, tree-like picture \cite{Corominas-Murtra:2011}.

\subsection{Feedfordwardness}

In the ${\cal G}_{\cal C}$, since the elements within a $SCC$ cannot be intrinsically ordered, $SCC$s constitute non orderable 
modules of the feedforward structure. As Fig. (\ref{Fig:fig_1}h) shows, $SCC$s represent a violation of the order at a given point of the feedforward 
condensed structure. Here, the size of the $SCC$s, but also 
their position in the feedforward structure are key elements for the quantification of the impact of cyclic modules in the feedforward structure, since the 
higher the $SCC$ position, the larger the number of its downstream dependencies. According to this, we define feedforwardness ($F$, being $0 \le F \le 1$), a measure that weights the impact of cyclic modules on the feedforward structure of the graph, where 
cyclic modules closer to the {\em top} of ${\cal G}$ will introduce 
a larger penalty on hierarchical order than those placed at the bottom. For every path $\pi_k$ 
starting from the top of ${\cal G}_{\cal C}$ we compute the fraction of the nodes that it contains against the actual nodes of ${\cal G}$ it represents. 
Formally, 
if $v(\pi_k)$ is the set of nodes participating in the path $\pi_k$,
\[
F({\pi_k})=\frac{|v(\pi_k)|}{\sum_{v_i\in v(\pi_k)} \alpha_i},
\]
where, as illustrated in Fig. (\ref{Fig:fig_1}), $\alpha_i$ is the {\em weight} of node $v_i$ in the {\em node-
weighted} condensed graph. To obtain a statistical estimator of the impact of the location of cyclic modules within the causal flow described by the 
network, we first have to define the set $\Pi_M$, which is the set containing all possible paths starting from the set of maximal nodes, $M$, and ending 
in any other node of ${\cal G}_{\cal C}$. Now,  $F({\cal G})$ is, thus, simply the average of $F({\pi_k})$ over all elements of $\Pi_M$, i.e.:
\begin{equation}
F({\cal G})=\langle F\rangle_{\Pi_M}.
\label{eq:F}
\end{equation}

\subsection{Orderability}

As hierarchy is grounded on the concept of order, we need a descriptor that accounts for how orderable is the graph under study. Ranging from a fully 
cyclic graph to a feedforward structure {\em orderability}  ($O$) lies in the range $0 \le O \le 1$ and it is defined as the fraction of the nodes of the 
graph ${\cal G}$ that {\em does not} belong to any cycle.  These nodes, therefore, make part of  the fraction of the network that can be actually ordered. 
Such an orderable fraction provides a raw but very meaningful estimator about how ordered is the set of nodes within the graph (see Figs. (\ref
{Fig:fig_1}i-l)). Formally, we define $O({\cal G})$ as:
\begin{equation}
O({\cal G})=\frac{|\{ v_i\in V_{\cal C}\bigcap V\}|}{|V|}.
\label{eq:o}
\end{equation}

Once we exposed the formal description of the different hierarchy indicators, i.e. treeness ($T$), feedforwardness ($F$) and orderability ($O$),  we proceed to collect them together to create a space $\Omega$ where networks can 
be properly evaluated and compared.

\begin{figure*}
\begin{center}
\includegraphics[width=16.5cm]{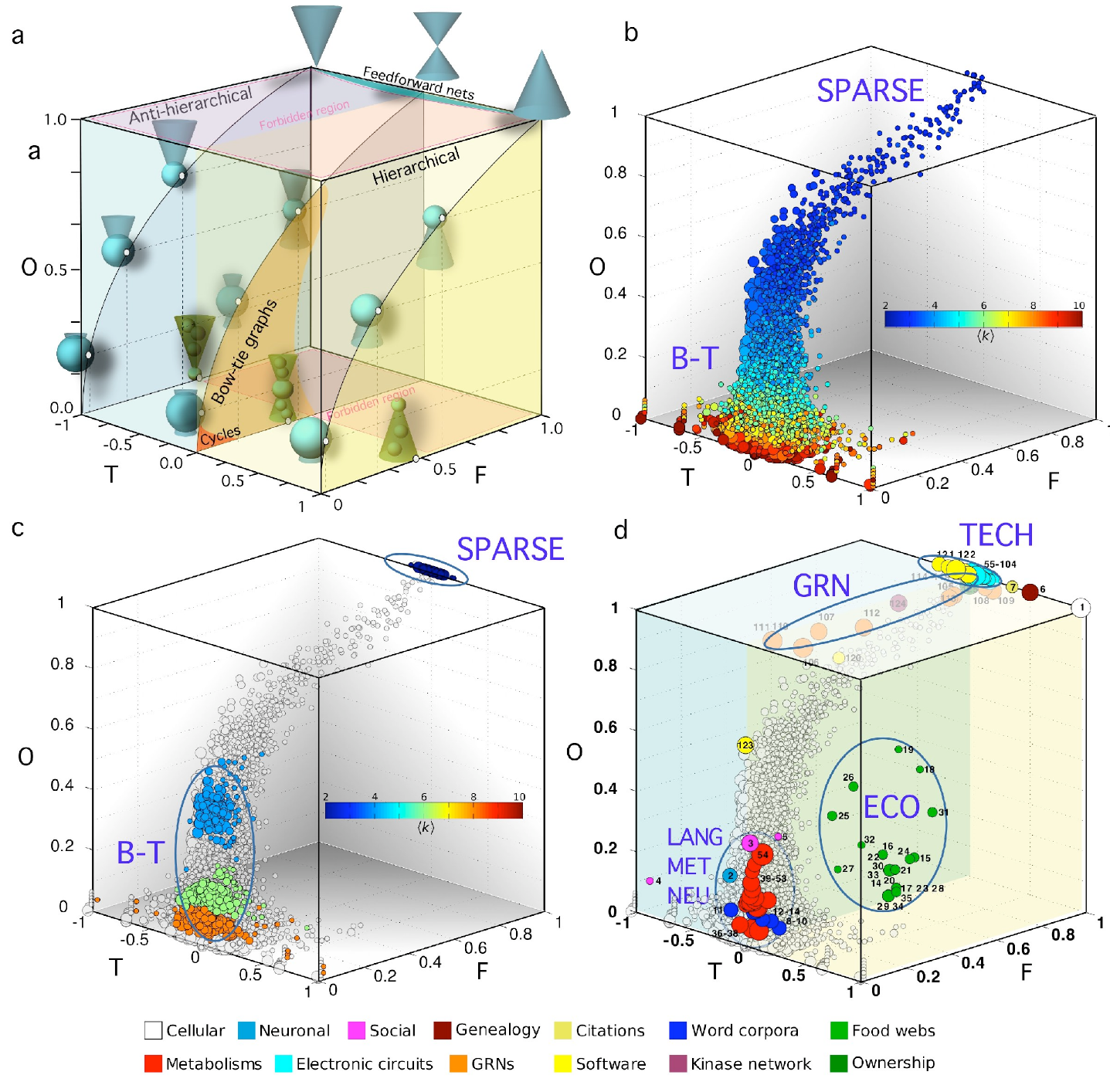}
\end{center}
\caption{The morphospace of possible hierarchies $\Omega$. (a) Different morphologies and their respective location within $\Omega$ (see Fig (1)). 
Green icons represent unlikely configurations. (b-c) The occupation of $\Omega$ by an ensemble of random models. 
This set includes Erd\"os-R\'enyi (ER) graphs with different sizes ($100$, $250$, $500$) and average degrees $\langle k \rangle$ (see color bar).  
Symbols are proportional to network size. (c) Morphospace occupation of the Callaway growing network model overlapped with the ER ensemble as a 
reference. Three network sizes  ($100$, $250$, $500$) and four connectivities ($\langle k \rangle=2,4,6,8$) are present (see also section \ref{Model networks} of the Appendix for further models 
and other details). (d) The coordinates of the $125$ real networks, colored and sized according to type and number of nodes, respectively. Cellular (\cite{Goni:2010} adapted from  WormBase.org),  Neuronal (\cite{White:1986} taken from Newman's dataset collection). Social, Genealogy, Citations and Ownership were taken from Pajek database (Vladimir Batagelj and Andrej Mrvar ${\rm http://pajek.imfm.si/doku.php?id=data:index}$). Word corpora  were generated according \cite{Ferrer:2001}. Food webs  were taken Pajek dataset from Ulanowicz colecction. See  references for Metabolisms \cite{Jeong:2000,Ma:2003,Ma:2007}, Electronic circuits \cite{Ferrer:2001}, GRNs \cite{Rodriguez-Caso:2009,Nitin:2010}, Software \cite{Valverde:2005} and Kinase networks \cite{Nitin:2010}. Network data are available upon request. See section  \ref{Sec:Network dataset} of the Appendix for more details about data compilation.  Numbers and colors 
are network identifiers in SI2. GRN, MET, LANG, NEU, ECO and TECH stand for Gene Regulatory, Metabolic, Linguistic, Neuronal, Ecological and 
Technological networks.}
\label{Fig:fig_2}
\end{figure*}

\section{The definition of the Morphospace $\Omega$}
Once our coordinates are defined, let us first see what is the repertoire of possible patterns that could be observed. According to our formalism, the hierarchical features of any directed network are given by a point $\mathbf{u}({\cal G})$ in a 3D morphospace  
$\Omega$, being $\Omega\subset [-1,1]\times[0,1]\times[0,1]$ (Fig. (\ref{Fig:fig_2}a) and Fig (\ref{fig:toys}) of the Appendix). The point $\mathbf{u}({\cal G})$ represents 
the graph ${\cal G}$ by three coordinates, 
\begin{equation}
\mathbf{u}({\cal G})=(T({\cal G}),F({\cal G}),O({\cal G})),
\label{eq:u}
\end{equation}
in the morphospace $\Omega$, according to its hierarchical properties.
Using the schematic representation of graphs outlined in Fig. (\ref{Fig:fig_1}), we define an intuitive icon associated to each kind of graph. As 
summarized in Fig. (\ref{Fig:fig_2}a), the perfect hierarchy is located at $\mathbf{u}({\cal G})=(1,1,1)$, whereas the completely non-hierarchical system -
a totally cyclic network- is located at $\mathbf{u}({\cal G})=(0,0,0)$. Interestingly, orderability and feedforwardness provide complementary information 
that defines  forbidden regions of the morphospace. Since $F({\cal G})=1$ is only possible  when $ O({\cal G})=1$, feedforward networks belong to the 
$F({\cal G})=O({\cal G})=1$ line. Given $O({\cal G})=1$, not other $F({\cal G})\neq 1$ is permitted by definition. Attending to $F$ and $O$, we find an 
interesting region, defined  by $O({\cal G})=0$ and $F({\cal G})>0$. It is worth to stress that, while $O({\cal G})$ and $F
({\cal G})$ converge in their upper bound  in a single value that defines the region of feedforward networks, they differ when $O$ goes to zero. This is 
because $F({\cal G})$ deals with the condensed graph while including \textit{condensed} modules but $O({\cal G})$ is about the nodes out of cycles. 
This little difference allows to unravel a family of rare networks highlighted by its extreme configuration -as shown by green icons in Fig. (\ref{Fig:fig_1}a). Essentially they are formed by chains of small $SCC$s disposed in a feedforward structure.

From these two axes accounting for the cyclic nature of networks, the coordinate $T$ provides additional information about the organization 
of the resulting feedforward structure after network condensation. Attending to the concept of a pyramidal structure, the plane separating hierarchy from 
anti-hierarchy ($T=0$) defines a family of symmetric structures, where the downstream path diversity 
is canceled by the uncertainty resulted when reversing these paths. Interestingly, when cycles are incorporated to this structure at $T=0$, the resulting 
structure is a bow-tie organization. In this particular structure, a large $SCC$ occupies a central position in a quite symmetrized feedforward 
structure of inputs and outputs \cite{Broder2000}. As shown in Fig. (\ref{Fig:fig_2}a), the larger  the $SCC$, the lower are the values of $F$ and $O$.

\section{Null models and real network analysis}

Since null models do not consider optimal designs or functional constraints, they provide good insights about the part of the morphospace where no 
selection pressure is at work. Fig. (\ref{Fig:fig_2}b-c) shows that both homogeneous and broad random networks 
occupy (basically) the same region in $\Omega$. This co-occupation, within the bow-tie plane with $T({\cal G})=0$, shows that 
random graphs appear located right in the middle between hierarchical and anti-hierarchical structures, 
independently of the type of degree distribution of the studied models. That implies, among other things, that 
they account for both bow-tie structures and feed-forward sparse webs. What about real nets? 
Here we use $N_n=125$ real networks encompassing $13$ classes of natural and artificial systems 
(see SI2 for numerical details of ${\mathbf u}({\cal G})$ values). As Fig. (\ref{Fig:fig_2}d) shows, a few 
isolated systems reach the boundaries of $\Omega$: a cell lineage located at ${\mathbf u}({\cal G})=(0.95,1,1)$ and 
a small social network located at ${\mathbf u}({\cal G})=(-1, 0.06, 0.09)$. However, most networks fall into four clusters. 

\begin{figure}
\begin{center}
\includegraphics[width=8.5cm]{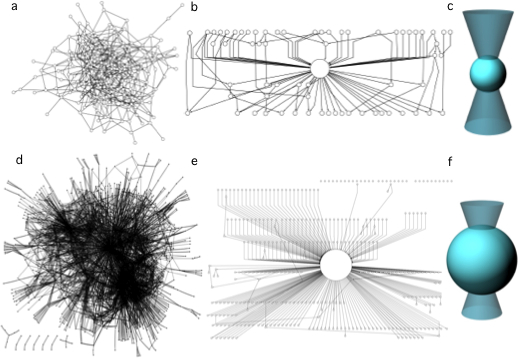}
\caption{Hierarchical order from fluctuations. Bow-tie networks have been described as obvious 
examples of optimized structures that pervade robust functional behavior in cellular and technological networks. However, 
a close analysis reveals that the underlying organization of random networks with a given average number of links already 
displays such kind of pattern. Here a random network (a) with $200$ nodes and an average degree of four links per node is built. 
The resulting ${\cal G}_{\cal C}$ (b) displays a very clear bow-tie organization (c). The human metabolic network (d), although non-homogeneous 
and resulting from evolution, displays a similar pattern (e-f). Both graphs are close within $\Omega$. Network layouts generated using Cytoscape software \cite{Smoot2011}.}
\label{Fig:fig_nova}
\end{center}
\end{figure}

First,  a group consisting of metabolic, neural, linguistic and some social networks is found at the lower part of the bow-tie domain, clearly embedded 
within the cloud of random graphs (Fig. \ref{Fig:fig_2}d) . Interestingly, randomized nets of this group 
 show a similar behavior although with a more central position within the 
cloud of random nets (see Fig. (\ref{Fig:fig_nova}). An interesting case 
is given by the presence of bow-tie patterns in metabolic networks \cite{Ma:2003b}.
They display a large central cycle, much larger than that observed in their randomized counterparts. This likely reflects 
the advantage of reusing and recycling molecules.
The second group placed at the $O({\cal G}) \approx 1$ plane shows a narrow band of  feedforward nets including 
electronic circuits with $-0.2<T({\cal G})<0.2$ and software graphs slightly biased to negative $T({\cal G})$ values. Here too the dispersal seems 
consistent with what is expected from very diluted random graphs (see Fig. (\ref{Fig:fig_2}c)). This sparseness is a consequence of engineering 
practices focused on reducing the wiring costs while keeping the system connected \cite{Ferrer:2001}.  

The third group  displays slightly positive values of $T({\cal G})$ and  is composed of 
graphs with cycles of small size but with a predominant position at the feedforward structure 
giving rise to a very high $O$ with variable $F({\cal G})$. These are gene regulatory networks plus a protein kinase network. $F({\cal G})$ dispersion is 
due to the variable size of modules located at the top of the structure. The special location in $\Omega$, far from the random cloud, is caused by a small fraction of  
genes, the DNA-binding elements (transcription factors), located at the top of the network, which participate in cycles. Finally, the fourth group is defined 
by an isolated cluster of ecological flow graphs, located around  ${\mathbf u}({\cal G}) = (0.35,0.45, 0.25)$. Their $T({\cal G})>0$ values indicate a 
certain degree of pyramidal structure and the low $O({\cal G})$'s are consistent with an important role played by loops. The special status of these networks (not shared 
by other webs) is consistent with the well known picture of a trophic pyramid combined with the presence of recycling \cite{Wickens1988, Allesina:2005}. 

The four clusters point to different scenarios pervading the origins of their hierarchical organization. One key observation is that most data sets are 
found within the envelope predicted by the ensembles of random graphs (see Fig. (\ref{Fig:fig_nova})). Since these 
null models do not consider optimal designs nor functional traits, we conclude that, as it was reported in the context of modularity \cite{Guimera2004, Sole:2008}, hierarchical order may be a byproduct of inevitable random fluctuations, which  spontaneously generate graph correlations, 
Bow-tie networks, which have been suggested to define a flexible, robust and evolved type of systems \cite{Kitano2004a, Stelling2004}, perhaps resulting from selection for robustness would be a byproduct of the generation rules responsible for network growth.

\section{Morphospace accessibility by driven evolution}

The previous results raise the question of how the voids in the morphospace need to be 
interpreted. In order to decide whether they are simply 
forbidden or have not been reached by evolution, we used an evolutionary search algorithm 
\cite{Marin:1999} which, starting from a random configuration, tries to get the points of a evenly 
gridded partition of the morphospace. The results provide a picture about how accessible are the 
different regions of $\Omega$. Detailed information about the evolutionary algorithm is found in section \ref{Sec:AccessibilityEvolution} of the
Appendix.  Very briefly, the evolutionary algorithm starts from a given set  of small random graphs 
${\mathbf u}  \in \Omega$
 belonging to the cloud of null models and networks. Given  a target point ${\mathbf u}^*$ of $\Omega$, these graphs are evolved by a random process of link addition and deletion with selection of networks minimizing the distance  
$||{\mathbf u}_t-{\mathbf u}^*||$. The number of nodes for every graph, $|V|=50$, remains constant in this process and graphs must belong to a single connected component. Only the number of links and their distribution  are affected by the evolutionary algorithm.  In 
this way the algorithm explores the network space, approaching the desired point and sometimes reaching it. The high computational  cost of this experiment makes difficult to operate with larger networks. However, it is worth to note that their small size provides an advantage for the evolutionary search in the change of the network configuration since few changes in the connections have by general a deep impact in their structure. In this way the use of small network sizes contribute to an efficient exploration, providing a coarse grained picture of network reachability. 

\begin{figure}
\begin{center}
\includegraphics[width=7cm]{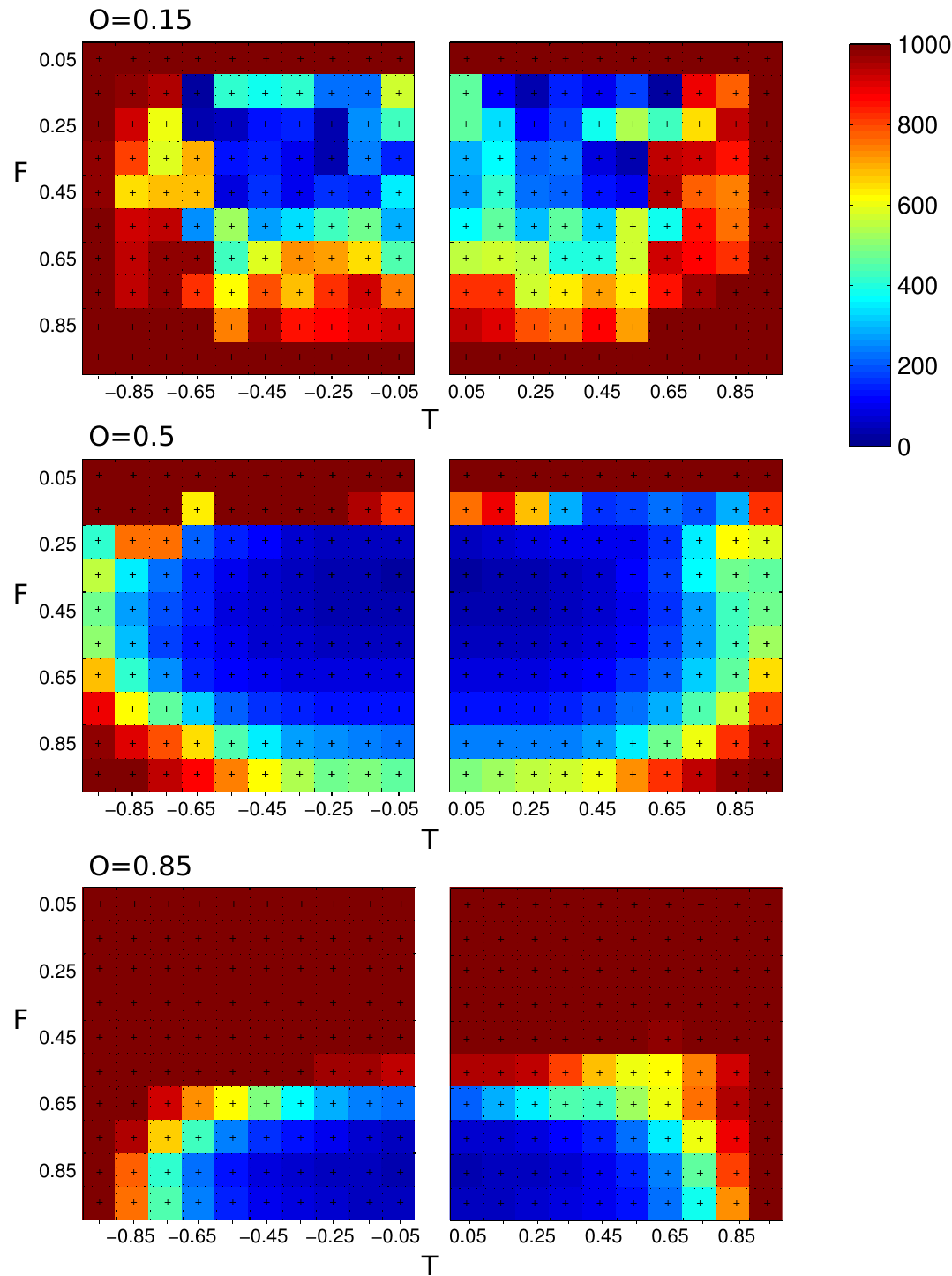}
\caption{Morphospace accessibility by driven evolution of random networks. Here three sections of the morphospace were explored by an 
evolutionary algorithm (see section \ref{Sec:AccessibilityEvolution} of the Appendix for algorithm definition). To study the reachability of the regions of $\Omega$, the space was evenly spaced in  300 
target points ${\mathbf u}^*$ distributed in a partition of $10\times10$ of the $TF$ plane and three values of $O({\cal G})=\{ 0.15, 0.5, 0.85\}$. Target 
points are labeled by crosses in the charts. Selection operated during $G=10^3$ 
generations. Color indicates at which generation were acquired every grid in an average 
of 250 evolutionary experiments. Blue squares indicate that they are -in average- accessible after a few rounds of the process whereas red ones are unreachable before $1,000$ iterations.}
\label{Fig:Accessibility}
\end{center}
\end{figure}

Fig. (\ref{Fig:Accessibility}) reveals that the cloud of null models  represented in Fig. (\ref{Fig:fig_2}b) is easily accessible, as indicated by the dark blue color of the region. As expected hierarchical ($T>0$) and anti-hierarchical regions ($T<0$) are quite symmetric. Deviations are due to the dispersion produced by the finite-size statistics, specially when the resulting ${\cal G}_{\cal C}$ become very little, by the imposition of a high number of cycles, as it happens  for $O=0.15$. This is the reason why, at $O=0.15$, reachability is rather heterogeneous. The solutions for this orderability are only possible by forcing to the network population to exhibit a large fraction of the nodes within cycles. This constraint inevitably produces a ${\cal G}_{\cal C}$ with just a handful of nodes. Then, as it happens for a broad number of topological measures, values of  $T$ and $F$ are extremely sensible to small
variations in network configurations. Such a trend is less dramatic when $O$ increases (see $O=0.5$ and $O=0.85$ of Fig. \ref{Fig:Accessibility}) since the fraction of nodes belonging to cycles is small enough to produce a rich combination of configurations in the resulting node-weighted condensed graphs.

However, the most interesting results here concerns the presence of inaccessible regions, labeled in dark red color. Low levels of $O$ seem to reduce the space of possible conformations. At $O=0.15$ the extremes of $T$ and $F$ are inaccessible under our \textit{in silico} evolutionary experiments. Such a behavior is relaxed at $O=0.5$ Here, a large region of high reachability  is observed for extreme values of $T$ but, not occupied by real networks. This may indicate that  a part of $\Omega$ is accessible and yet 
not occupied, suggesting that the spontaneous correlations created by random fluctuations provide the source of order for free. As a consequence, non-
adaptive processes would have played a major role in shaping hierarchies in nature \cite{Lynch2007}. 

Finally, an interesting trend is observed when $O$ approaches its upper bound. The larger $O$, the more reduced is the range of the \textit{possible} of $F$. Close to $O=1$,  there only exists graph configurations for $F=1$, coinciding with feedforward networks encompassing, electronic circuits and software networks. 

\section{Concluding remarks}

Both biological and cultural evolution operate under a number of deep constraints \cite{Sole2013}. Some of them 
result from the underlying rules of network growth and change, which strongly limit the repertoire of potential designs. 
An important question posed by evolutionary theory is the nature and relevance of such constraints in shaping the 
space of the possible. Our study provides a rationale for exploring the possible and the actual in complex networks under a 
static view dominated by causal relations among components and modules. In this context, the inclusion of functionality, dynamics or 
weighted structures has not being taken into account and should be the object for further work. By defining a general 
space of hierarchical webs, we are able to detect the presence of a rather limited domain occupied by real and random null-models.  
The large voids surrounding these clusters of webs (defining four major groups) are partially inaccessible and partially reachable, 
as shown by means of a directed evolution algorithm. The majority of webs display a balance between 
integration of multiple signals and control over multiple targets under a bow-tie structural pattern. The computational nature of 
regulatory networks and the combination of layers and cycles common to energy flows in food webs separate them from 
this large cluster. The matching of random and real webs in the first two clusters suggests that their hierarchical features can be accounted for from the spontaneous 
correlations associated to random graphs of a given degree, indicating that the observed webs are simply the most probable ones. 
By connecting network theory with theoretical morphology a powerful picture of complexity emerges, which allows us to both 
characterize hierarchical order and provide an evolutionary framework to explain how hierarchy emerges in nature. 
The formalism presented in this work provides a suitable framework for the quantitative approximation for the study  of hierarchical organizations, and links to non-equilibrium thermodynamics could be defined in the future, attending the similarity of certain approaches \cite{Rinaldo:1996, Wissner-Gross:2013}.  Further effort in the inclusion of the strength of relations among elements from empiric data as weighted graphs would contribute for a more accurate view of the hierarchy of systems. Future work in the development of generative models for the study of the emergence of  hierarchy will be of strong interest in the study of dynamics in the exploration of the limits of what is possible for natural, technological and social organizations. 
 
\vspace{0.5 cm}
{\bf Acknowledgements} We would like to thank the members of the Lab for useful discussions. RVS thanks 
D. Erwin, E. Smith, G. West and M. Gell-Mann for useful discussions on hierarchy. We thank Olaf Sporns for useful comments.  
We thank Wormbase, Vladimir Batagelj and  Andrej Mrvar  for Pajek dataset and Mark Newman for his network dataset.
This work has been supported by grants of the James McDonnell Foundation, 
the Bot\'in Foundation and by the Santa Fe Institute.


\newpage

\appendix

\section{Structure of  the Appendix}
This Appendix presents in a self-contained way the conceptual issues that lead us to the rigorous formalization of hierarchy in complex networks. 
Although some of these concepts can be found in standard handbooks on graph theory -for example \cite{Gross:1998}-, we present all them from scratch  in order to provide the reader with a consistent and clear mathematical apparatus. The aim is to remove any inconsistency in notation when introducing original concepts. The type of object studied here is any kind of {\em connected} directed graph, i.e., any directed graph composed by a single component. Although the formalism could be applied over graph structures consisting in more than a single component -as usually happens when working with random graphs- we assume that unconnected components have no causal relation of any kind and, therefore, the concept of hierarchy looses any meaning. As we shall see, the possibility to relate parts of the system by means of the topological information is basic for our definition of hierarchy. 

The appendix is structured as follows: First we present in an axiomatic way the features a given system based on relations among objects must hold in order to be considered {\em perfectly hierarchical}. Deviations from this perfect configuration have to be properly quantified, giving rise to the three coordinates of the hierarchy. Then, we provide the basis to work with, revising several basic definitions of directed graphs and some of their most salient properties. We focus our attention on the condensation operation and on the layered structure of the resulting condensed graph. Basic definitions are followed by the proper definition of the hierarchy coordinates. Such definitions are detailed and described with the aim of conveying the reader the flavor and intuitions that underlie them. Once the formal framework is properly described
we explore the regions of the defined morphospace occupied by model networks. This section is followed by the list of studied networks and the randomization methods we used to explore the relevance of the observed results. It is important to stress that the rigorous and systematic confrontation of real data with their randomized counterparts elucidates much of the possible origins of the observed patterns. Finally, we explore the space of possible hierarchical configurations through an evolutionary algorithm whose rules are described in detail.

\section{The fundamentals of hierarchy (Postulates of Hierarchy)}
\label{Sec:Postulates}
The {\em perfectly hierarchical system} will have the following three properties:
\begin{itemize}
\item
{\em order}, 
\item
{\em reversibility} and 
\item
{\em pyramidal structure}.
\end{itemize}
Let us be more precise. Suppose we have a set $A=\{a_1,...,a_n\}$ and a set relation $R\subset A\times A$. Let $T(R)$ be the {\em transitive closure of $R$} \footnote{Given a set $A$, The transitive closure of a relation $R\subseteq A\times A$ is the minimal transitive relation $R'$ such that $R\subseteq R'$. See \cite{Kelley:1975, Suppes:1960} or \cite{Gross:1998}.}, we will say that $R$ {\em is hierarchical} if $R$ has the following properties:
\begin{itemize}
\item
{\em order},
\[
{\rm if} \;\langle a_i,a_k\rangle\in T(R ) , \;{\rm then} \;\langle a_k,a_i\rangle\notin T(R )
\]
({\em There are no cyclic relationships: we can {\bf order} the elements \footnote{In this case, we impose that $T(R )$ is a {\em strict partial order}. $K\in A\times A$ is a strict partial order if it is i) $(\forall a_i\in A) (\langle a_i, a_i\rangle\notin K)$ ({\em non-reflexive}) ii) $(\langle a_i,a_k\rangle\in K)\Rightarrow(\langle a_k,a_i\rangle\notin K$) ({\em antysimetric}) and iii) $[(\langle a_i,a_k\rangle\in K)\wedge (\langle a_k,a_j\rangle\in K)]\Rightarrow(\langle a_i,a_j\rangle\in K)$ ({\em transitive}).})}
\item
{\em reversibility}
\[
{\rm if} \;\langle a_i,a_k\rangle\in R \;{\rm and}\; \langle a_j,a_k\rangle\in R,\; {\rm then}\; a_i=a_j
\]
({\em There is only one commander for any commanded: The chain of commands is {\bf reversible}}.)
\item
{\em pyramidal structure}

{\bf a}) ${\rm if} \;\langle a_i,a_k\rangle\in R \;{\rm then}, \; \exists a_{j}\neq a_k\;{\rm such\;that}\; \langle a_i,a_j\rangle\in R $
 ({\em A commander commands more than a single element.}) 

{\bf b}) $(\exists ! a_k\in A)\;{\rm such\;that}\;(\forall a_i\in A\setminus \{a_k\})(\langle a_k,a_i\rangle\in T(R ))$ 
({\em There is only a single element which is not commanded by another element.}) 

{\bf c}) Let $J_i$ be the set of commanders of element $a_i$, i.e.: $J_i=\{a_k\in A: \langle a_k, a_i\rangle \in T(R )\}$
If $a_i, a_j\in A$ are such that $\nexists a_k\in A$ such that $\langle a_i, a_k\rangle\in R$ or $\langle a_j,a_k\rangle\in R$, then
$|J_i|=|J_j|.$
({\em All the elements of the bottom are subjected to a chain of commands of the same length. These last properties give the 
{\bf pyramidal structure} of $R$})
\end{itemize}

\begin{figure}[h]
\begin{center}
\includegraphics[width=7cm]{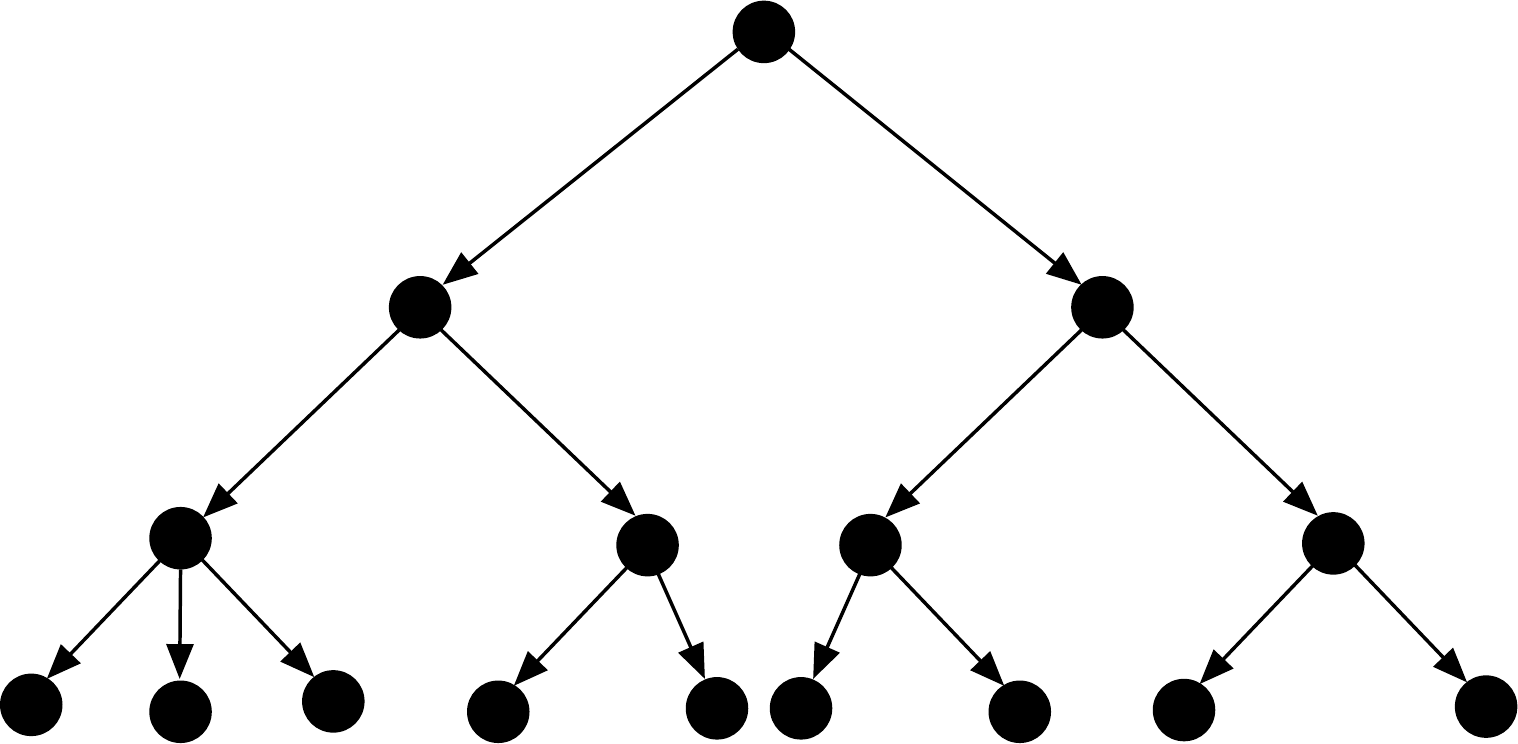}
\caption{A possible chart of relations depicting a perfectly hierarchical system.}
\label{fig:Postulates}
\end{center}
\end{figure}

It is straightforward to realize that the directed, inverted {\em tree} in which all the arrows go {\em downwards} starting from a single root naturally emerges as the graphical description of the perfect hierarchical system -see fig. (\ref{fig:Postulates}). The rigorous proof of this statement is given as a lemma  in section \ref{Sec:DAGs} when the appropriate formalism is developed.

{\bf The measure of hierarchy must properly quantify the deviations from this perfect structure, considering these three postulated properties.}

\section{Graph definitions}

The whole set of relationships between elements of the systems under scrutiny define a {\em directed graph}. In any directed graph, links describe some kind of causal relation between two elements of the set which is considered relevant for the system.
We thus begin revising several facts about directed graphs. These basic notions can be found in any standard textbook on graph theory, like \cite{Gross:1998}.

\subsection{Directed Graphs}
\label{Sec:DirectedGraphs}
Let    ${\cal    G}(V,    E)$    be   a    directed    graph,    being
$V=\{v_1,...,v_n\},\;|V|=n$, the  set of nodes,  and $E=\{\langle v_k,
v_i\rangle, ..., \langle v_j, v_l\rangle\}$ the set of arcs -where the
order, $\langle v_k,  v_i\rangle$ implies that there is  an arc in the
following direction:  $v_k\rightarrow v_i$.  Given a  node $v_i\in V$,
the  number of  outgoing links,  to be  written as  $k_{out}(v_i)$, is
called the {\em  out-degree} of $v_i$, and the  number of ingoing links
of  $v_i$  is  called  the   {\em  in-degree}  of  $v_i$,  written  as
$k_{in}(v_i)$. The {\em adjacency matrix} of a given graph ${\cal G}$,
$\mathbf{A}({\cal   G})$   is   a $n\times n$ matrix where   $A_{ij}({\cal   G})=   1
\leftrightarrow \langle v_i,  v_j\rangle\in E$; and $A_{ij}({\cal G})=
0$ otherwise.   Through the  adjacency matrix, $k_{in}$  and $k_{out}$
are computed as
\begin{equation}
k_{in}(v_i)=\sum_{j\leq n}A_{ji}({\cal  G});\;\;\;\;k_{out}(v_i)=\sum_{j\leq n}A_{ij}({\cal G}).
\label{kin}
\end{equation}

A {\em path} from node $v_i\in V$ to node $v_j\in V$, $\pi_k(v_i, v_j)$ in a directed graph is an alternated sequence of nodes and links:
\[
\pi_k(v_i, v_j)=v_i,\langle v_i, v_j\rangle, v_j,...,v_{\ell},\langle v_{\ell},v_j\rangle, v_j
\]
such that
\[
\langle v_i, v_j\rangle,...,\langle v_{\ell},v_j\rangle\in E.
\]
(The $k$ subscript is due to the possible presence of more than a single path from $v_i$ to $v_j$). The {\em length} of a path, $\ell(\pi_k(v_i, v_j))$ is the number of edges appearing in the sequence. We observe that some edges can appear twice or more in the sequence. In that case, we take into account all appearances of the edge.
Since a path $\pi_k(v_i, v_j)$ is an alternating sequence of nodes and links, it seems natural to define two sets:  $v(\pi_k(v_i, v_j))$ as  the set of all nodes present in the path and $e(\pi_k(v_i, v_j))$,  the set of all edges present in the path.  Then we define the set $\Pi({\cal G})$ as the (possibly infinite) set of all paths that can be defined in ${\cal G}$. 

A {\em cycle} in a directed graph is the subgraph formed by the edges and nodes defining a path which begins and ends in the same node, i.e, if
\[
v_i,\langle v_i, v_j\rangle, v_j,...,v_{\ell},\langle v_{\ell},v_i\rangle, v_i
\]
such that $\langle v_i, v_j\rangle,...,\langle v_{\ell},v_i\rangle\in E$. The set of cycles of a given graph is ${\cal C}({\cal G})=\{C_1,...,C_k\}$. If $C_k\in {\cal C}({\cal G})$ is maximal, i.e.,
\[
\nexists C_i\in {\cal C}({\cal G}): C_k\subset C_i,
\]
then, $C_k$ is called a {\em Strongly Connected Component}, hereafter $SCC$. 
We will refer to the set of $SCC$'s as $\Sigma ({\cal G})=\{S_1,...,S_{\ell}\}$. Since every $SCC$ is itself a subgraph, we can refer to it as $S_k=S_k(V_{S_k},E_{S_k})$, where $V_{S_k}$, $E_{S_k}$ are the nodes and the edges of $S_k$, respectively.

The {\em underlying graph} of a given directed graph ${\cal G}$, to be written as  ${\cal G}^u$, is the undirected  graph ${\cal G}^u(V,E^u)$
obtained by  substituting all arcs  of $E$, $\langle  v_i, v_k\rangle,
\langle  v_j, v_s\rangle,....$ by undirected edges giving  the set  $E^u=\{\{ v_i,
v_k\},  \{ v_j,  v_s\},...\}$. A  directed graph ${\cal G}$  is said  to be  {\em
  connected}  if for  any pair  of nodes  $v_i,  v_l\in V$
there is a finite,  undirected path linking them; i.e., a finite sequence 
\[  
  v_i,\{v_i, v_k\},v_k ,...,v_s,\{v_m,v_{\ell}\},v_{\ell}
\]
where  $\{v_i, v_k\},...,\{v_m,v_{\ell}\}\in E_u$. Notice that now the links are depicted by unordered pairs, therefore $\{v_i, v_k\}=\{v_k, v_i\}$.

Given a graph ${\cal G}(V,E)$, a {\em component} $\gamma_j$ is a maximal subgraph of ${\cal G}$ by which, for every pair of nodes $v_k, v_i$ belonging to it there is an undirected, finite path linking them. We refer to the set of components of ${\cal G}$ as 
\[
\Gamma({\cal G})=\{\gamma_1,...,\gamma_k\}. 
\]
It turns out that a connected graph has only one component, the graph itself. In general, we will work with connected graphs. It is worth to note that, connectedness may be lost during a randomization process but,  as we shall see in section \ref{Sec:RealNetworks}, the randomization methods we use respect the structure of components of the graph -i.e., connectedness if it be.

\subsection{Directed Acyclic Graphs}
\label{Sec:DAGs}
A  {\em  directed  acyclic graph} or {\em feedforward graph} -hereafter, $DAG$- is  a
directed graph characterized  by the absence of cycles. 
The first consequence of the absence of cycles is the non-existence of infinite paths within -a finite- ${\cal G}$.
$DAG$s have been also referred to as {\em ordered graphs} because it is always possible to {\em topologically sort} a $DAG$ \cite{Gross:1998}. A topological sorting consists in  numbering all the nodes $v_1,v_2,...,v_n$ in such a way that arcs always point to nodes having higher numerical label than their origin, i.e., for any pair $v_k,v_j$ such that $k>j$, then
\[
\langle v_k,v_j\rangle \notin E.
\]
This property will be crucial to justify the link between hierarchy, order and causal relation. This sorting implies that a $DAG$ can represent a more or less entangled structure of interconnected causal processes.

We stress in the analogy between order theory and $DAG$ structures by defining following set:
\begin{equation}
M=\{v_i\in V:k_{in}(v_i)=0\},
\label{MaximalSet}
\end{equation}
 to be  named the set  of {\em maximal  nodes} of ${\cal G}$.  Complementarily, one can define the set of nodes $\mu$ as
\begin{equation}
\mu=\{v_i\in V:k_{out}(v_i)=0\},
\label{minimalSet}
\end{equation}
which will be referred to as the set of {\em minimal nodes} of ${\cal G}$.

The set  of all  paths $\pi_1,...,\pi_s$, from $M$  to $ \mu$ is indicated as $\Pi_{M\mu}({\cal G})$.  Given
a node $v_i\in \mu$, the set of all paths from $M$ to $v_i$ is written
as  
\begin{equation}
\Pi_{M\mu}(v_i)\subseteq \Pi_{M\mu}({\cal G}).
\label{PiMmu}
\end{equation}
We observe that  the length of the {\em longest} path of  a finite $DAG$ is always bounded. We will refer to this numerical value as  $L({\cal G})$ which, taking into account the properties of the powers of the adjacency matrix, can be straightforwardly obtained \footnote{This contrasts with the case of a graph containing cycles, where, even in the case of being finite, cycles allow the presence of paths of infinite length.}:
\begin{equation}
L({\cal    G})=\max\{k:(\exists   v_i,    v_j\in   V:(\mathbf{A}^k({\cal  G}))_{ij}\neq 0)\}.
\label{K}
\end{equation}

A special type of $DAG$ is to be defined now, since, as we will show below, depicted the perfect hierarchy. A {\em directed tree}  in which {\em all the leafs have the same length} is a $DAG$ with $|M({\cal G})|=1$ in which all elements but the one in $M$ have $k_{in}=1$, either $k_{out}=0$ or $k_{out}\geq 2$ and in which $\forall v_i,v_j\in \mu ({\cal G})$, if $v_k\in M$, then $\ell(\pi(v_k,v_i))=\ell(\pi(v_k,v_j))$ -the {\em leafs} are all paths going from $M$ to $\mu$. See figure (\ref{fig:Postulates}) for an example of this kind of graph. 

Let us now take the set-theoretic framework used at the beginning of  this document and replace $A$ by $V$ (the set of nodes) and $R$ by $E\in V\times V$ (the set of links). Now we have a graph representation of the kind of relations that display a perfect hierarchical organization. It turns out that the type of graph defined above satisfies all the postulates defined in \ref{Sec:Postulates} is the graphical representation of a perfect hierarchical system. Let us state is as a formal lemma:

\noindent{\bf Lemma}:
{\em Under the hallmark described (see Postulates, section \ref{Sec:Postulates}), the kind of graph ${\cal G}$ depicting the perfect hierarchy is a directed tree  in which all the leafs have the same length. }

\noindent{\bf Proof}:
By the {\em order} property, ${\cal G}$ has no cycles. By the {\em reversibility} property the graph nodes of the graph have either $k_{in}=0$ or $k_{in}=1$. By the {\em pyramidal} property {\bf a)} the nodes have either $k_{out}=0$ or $k_{out}\geq 2$. By  {\em pyramidal} property {\bf b)} the graph is connected and $|M({\cal G})|=1$ and by  {\em pyramidal} property {\bf c)}, if  $v_k\in M$, then $\forall v_i,v_j\in \mu ({\cal G})$, then $\ell(\pi(v_k,v_i))=\ell(\pi(v_k,v_j))$ (notice that there is only one path to go from $v_k\in M({\cal G})$ to a given $v_i\in \mu({\cal G})$). Therefore, ${\cal G}$ is {\em directed tree}  in which {\em all the leafs have the same length}.

Now that we have the mathematical characterization of the perfectly hierarchical graph, we restart the conceptual presentation that will help us to properly evaluate deviations from the ideal graph.

\subsubsection{Condensation: Obtaining a $DAG$ from any directed graph}

Now we consider again the wide class of directed graphs, not only $DAG$s. Let $\Sigma({\cal G})=\{S_1,...,S_k\}$ be the set of $SCC$s. We will build the {\em condensed} graph of ${\cal G}$, to be referred to as ${\cal G}_{\cal C}$, in which every $SCC$ is merged into a single node, maintaining the links that connect a 
node of such $SCC$ with nodes out of such $SCC$. More formally, 
\[
{V}_{\cal C}=\Sigma({\cal G}) \bigcup \left[V\setminus \left(\bigcup_{\Sigma({\cal G})} V_{S_j}   \right)\right],
\]
i.e., nodes of ${\cal G}_{\cal C}$ are either $SCC$s of ${\cal G}$ or nodes that do not belong to any  $SCC$ of ${\cal G}$. Consistently, edges on ${E}_{\cal C}$ connect nodes of ${V}_{\cal C}$ in such a way that
\begin{widetext}
\[
(\forall v_i,v_j\in {V}_{\cal C})\langle v_i,v_j\rangle\in {E}_{\cal C} \Leftrightarrow\left\{
\begin{array}{ll}
\langle v_i, v_j\rangle\in E \\
(v_i=S_k\in\Sigma({\cal G}), v_j\in V)\wedge(\exists v_{\ell}\in S_k):(\langle v_{\ell}, v_j\rangle \in E)\\
(v_i=S_k\in\Sigma({\cal G}), v_j=S_{\ell}\in \Sigma({\cal G}))\wedge(\exists v_{m}\in S_k,\exists v_s\in S_{\ell}):(\langle v_m,v_s\rangle\in E);\;\;{\rm for} \;\;k\neq \ell.
\end{array}
\right.
\]
\end{widetext}
It turns out that, by definition, ${\cal G}_{\cal C}$ is a $DAG$.

Below we have an example of the condensation operation, step by step. We have a graph ${\cal G}$ (left). We then identify the $SCC$s of ${\cal G}$ (center) and collapse all nodes belonging to a $SCC$ in a single node, thus obtaining ${\cal G}_{\cal C}$ (right):
\begin{figure}[h]
\begin{center}
\includegraphics[width=8cm]{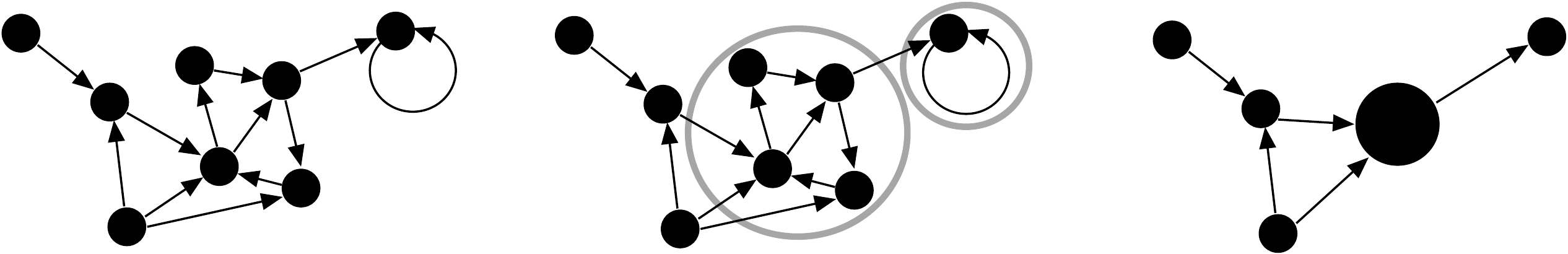}
\caption{The three stages of the condensation algorithm.}
\label{fig:Condenstion}
\end{center}
\end{figure}

Notice, as illustrated in figure (\ref{fig:Condenstion}) that ${\cal G}_{\cal C}$ is a $DAG$, as expected.

\subsubsection{{\em Node-weighted} condensed graph}

A conceptual step beyond the condensed graph is the {\em node-weighted} condensed graph.
All the computations presented in this paper will be performed over such a graph.  
Let us formally describe it:
In this graph, a weight $\alpha_i$ is assigned to every node $v_i\in V_{\cal C}$ in the following way:
\[
\alpha_i=\left\{\begin{array}{ll}
1\Leftrightarrow v_i\in V\bigcap V_{\cal C}\\
|S_i|\Leftrightarrow v_i\in \Sigma({\cal G}).
\end{array}
\right.
\nonumber
\]
Therefore, the sequence of 
\[
\Lambda=\alpha_1,..., \alpha_{|V_{\cal C}|}
\]
will be the sequence of weights of ${\cal G}_{\cal C}$. Clearly,
\[
\sum_{i\leq |V_{\cal C}|}\alpha_i=|V|.
\]
In plain words, $\alpha_i$ accounts for the number of nodes belonging to $V$ represented by a single node in ${V}_{\cal C}$. In the forthcoming example we detail the obtaining of such a graph from a given directed graph. We have the graph -left-; then, we identify its $SCC$s -center- and, then, -right- we label the nodes of ${\cal G}_{\cal C}$ with their corresponding $\alpha$'s:
\begin{figure}[h]
\begin{center}
\includegraphics[width=8cm]{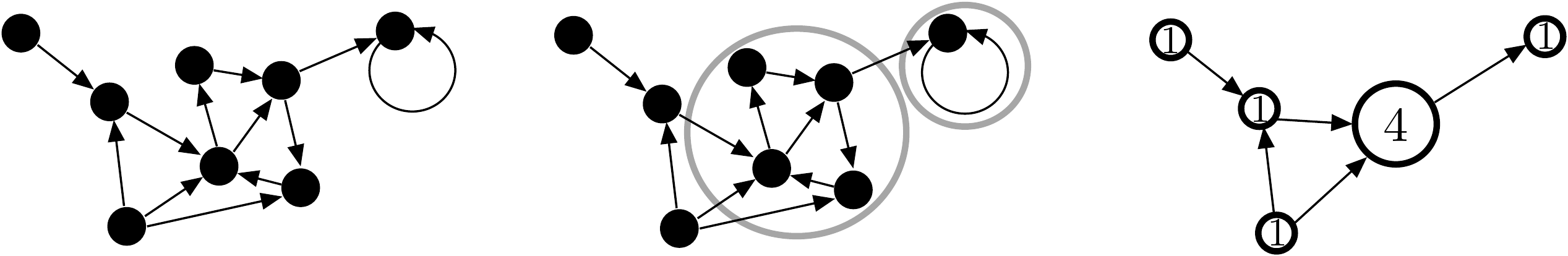}
\caption{Obtaining the {\em node-weighted} condensed graph ${\cal G}$.}
\label{fig:Alfes}
\end{center}
\end{figure}

\subsubsection{Dissection of the layers of the graph}
\label{Sec:Dissection}

The objective of this section is to rigorously define the dissection in layers of a given $DAG$. The non-cyclic nature of such graph structures enables us to define a finite set of layers. Layers can be identified through a  {\em backward} or {\em bottom up leaf removal} ($LR_b$) algorithm  or through a {\em forward} or {\em top down leaf removal} ($LR_f$) algorithm \footnote{A $LR_f$ algorithm works as follows: Given a $DAG$ ${\cal G}$, at every iteration we remove the nodes having $k_{in}=0$ until there is no node to remove. The set of nodes removed at every iteration define a layer of the $DAG$. It is straightforward that the number of steps needed is $L({\cal G})+1$ and so is the number of layers. A $LR_b$ algorithm works exactly in the same way, but removing nodes having $k_{out}=0$. The interested reader can go to \cite{Lagomarsino:2005, Rodriguez-Caso:2009}. }.
Although  $LR_f$ and $LR_b$ algorithms identify the same number of layers, the specific composition of every layer is not necessarily the same.
In addition, using these two algorithms, we can build two sets of subgraphs by successively removing layers of the $DAG$. As we shall see, these two sets will be very useful to derive the proposed measure of hierarchy.

Accordingly, we define two  partitions of ${V}_{\cal C}$,
$W=\{\omega_1,...,\omega_m\}$           and
$\tilde{W}=\{\tilde{\omega}_1,...,\tilde{\omega}_m\}$. The number of layers relates to $L({\cal G}_{\cal C})$ -see equation (\ref{K})- as follows:
\[
|W|=|\tilde{W}|=L({\cal G}_{\cal C})+1.  
\]
The members of
such partitions are the nodes defining the layers of the $DAG$,  computed by either  a $LR_b$ or $LR_f$  
algorithm -depending on which partition we generate, either $W$ or $\tilde{W}$, respectively.
Specifically,   using the bottom up approach -i.e., using a $LR_b$ algorithm- the first member of such partition is defined as the following subset of nodes:
\[
\omega_1=\{v_i\in V_{\cal C}:k_{out}(v_i)=0\}
\]
and, using the top-down approach -i.e., using a $LR_f$ algorithm- the first member is defined as, 
\[
\tilde{\omega}_1=\{v_i\in V_{\cal C}:k_{in}(v_i)=0\}.
\]
Clearly, $\omega_1=\mu$ and $\tilde{\omega}_1=M$.
With  the  above  subsets of  $V_{\cal C}$  we  can  define the  graphs  ${\cal
  G}_1(V_1, E_1)$, and $\tilde{\cal G}_1(\tilde{V}_1, \tilde{E}_1)$ in
the following way:
\[
V_1=V_{\cal C}\setminus\omega_1;\;\;E_1=E_{\cal C}\setminus \{\langle v_i,v_k\rangle:v_k\in \omega_1\}
\]
and
\[
\tilde{V}_1=V_{\cal C}\setminus\tilde{\omega}_1;\;\;\tilde{E}_1=E_{\cal C}\setminus
\{\langle v_i,v_k\rangle:v_i\in \tilde{\omega}_1\} .
\]
respectively.  Similarly, we build $\omega_2$ and $\tilde{\omega}_2$ as:
\begin{eqnarray}
\omega_2&=&\{v_i\in V_1:k_{out}(v_i)=0\},\nonumber\\
\tilde{\omega}_2&=&\{v_i\in\tilde{V}_1:k_{in}(v_i)=0\}.\nonumber
\end{eqnarray}
which, in turn, enables us to derive $V_2$, $E_2$, $\tilde{V}_2$ and $\tilde{E}_2$:
\begin{eqnarray}
V_2=V_1\setminus\omega_2;&&E_2=E_1\setminus \{\langle v_i,v_k\rangle:v_k\in \omega_2\}\nonumber\\
\tilde{V}_2=V_1\setminus\tilde{\omega}_2;&&\tilde{E}_2=E_1\setminus\{\langle v_i,v_k\rangle:v_i\in \tilde{\omega}_2\}.\nonumber
\end{eqnarray}
In the general case, if $V_0=\tilde{V}_0= V_{\cal C}$,
\begin{eqnarray}
V_{\ell}=V_{\ell-1}\setminus\omega_{\ell};&&E_{\ell}=E_{\ell-1}\setminus \{\langle v_i,v_k\rangle:v_k\in \omega_{\ell}\}.\nonumber\\
\tilde{V}_{\ell}=V_{\ell-1}\setminus\tilde{\omega}_{\ell};&&\tilde{E}_{\ell}=\tilde{E}_{\ell -1}\setminus\{\langle v_i,v_k\rangle:v_i\in \tilde{\omega}_{\ell}\},\nonumber
\end{eqnarray}
where
\[
1\leq\ell\leq L({\cal G}_{\cal C}).
\]
It is worth to note that
\[
E_{L({\cal G}_{\cal C})}=\tilde{E}_{L({\cal G}_{\cal C})}=\varnothing,
\]
and that
\[
V_{L({\cal G}_{\cal C})}\subseteq M;\;\;\;\tilde{V}_{L({\cal G}_{\cal C})}\subseteq \mu.
\]
The two previous sequences of subgraphs can be ordered by inclusion, namely
\[
 {\cal G}_{L({\cal G}_{\cal C})}\subset  ...\subset {\cal
  G}_1\subset {\cal G}_{\cal C},
\]
and
\[
 \tilde{\cal    G}_{L({\cal G}_{\cal C})}\subset
...\subset \tilde{\cal G}_1\subset{\cal G}_{\cal C}.
\]
The dissection of  ${\cal G}_{\cal C}$  in the above described two sequences of subgraphs will enable us to exhaustively explore the role of all layers in the further hierarchy measure. We finally define the set ${\cal W}({\cal G})$, containing the graph ${\cal G}_{\cal C}$ and all non-empty subgraphs obtained by means of the application of a leaf removal algorithm (either bottom up or top down) and which contain at least one link:
\begin{equation}
{\cal W}({\cal G})=\{{\cal G}_{\cal C}, \tilde{\cal    G}_{L({\cal G}_{\cal C})-1},. . ., \tilde{\cal G}_1, {\cal G}_{L({\cal G}_{\cal C})-1},. . . ,{\cal G}_1\}.
\label{eq:DefW}
\end{equation}
It is not difficult to check that
\[
|{\cal W}({\cal G})|=2L({\cal G}_{\cal C})-1.
\]

Let us provide an example in figure (\ref{fig:LeafRemoval}):  Starting from a $DAG$ (left), we identify the layers $\omega_1,...,\omega_5$  using a $LR_b$ algorithm (center), and the layers $\tilde{\omega}_1,...,\tilde{\omega}_5$ using a $LR_f$ algorithm   (right):
\begin{figure}[h]
\begin{center}
\includegraphics[width=8cm]{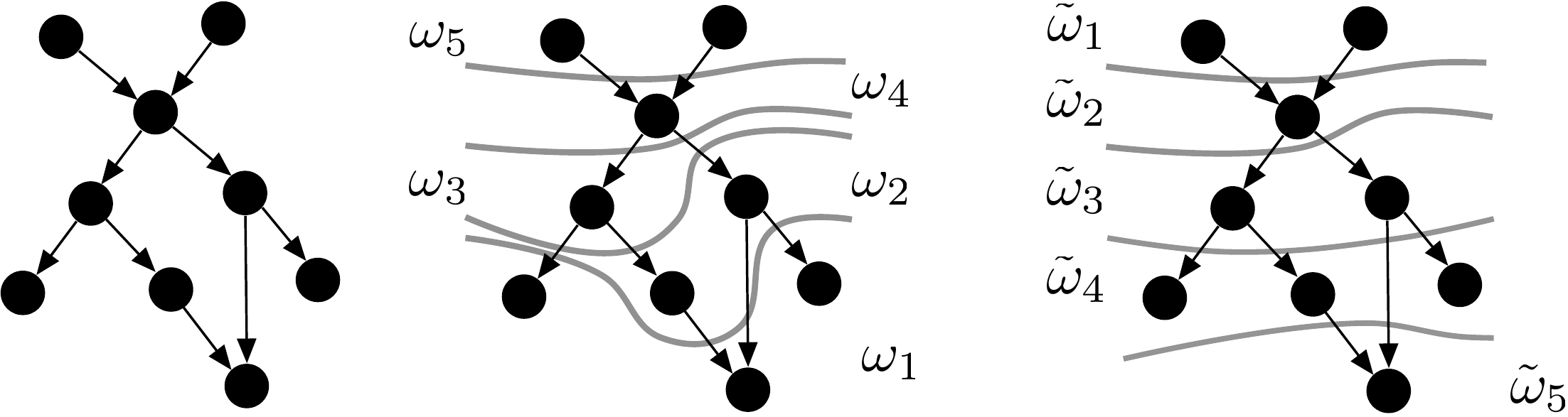}
\caption{The LR algorithms and the identification of layers. In this case, $|W|=5$ and $L({\cal G})=4$.}
\label{fig:LeafRemoval}
\end{center}
\end{figure}
Furthermore, we can obtain the sequence of graphs ordered by inclusion ${\cal G}, {\cal G}_1,...,{\cal G}_4 $ -notice that the last graph consists of two isolated nodes: 
\begin{figure}[h]
\begin{center}
\includegraphics[width=8cm]{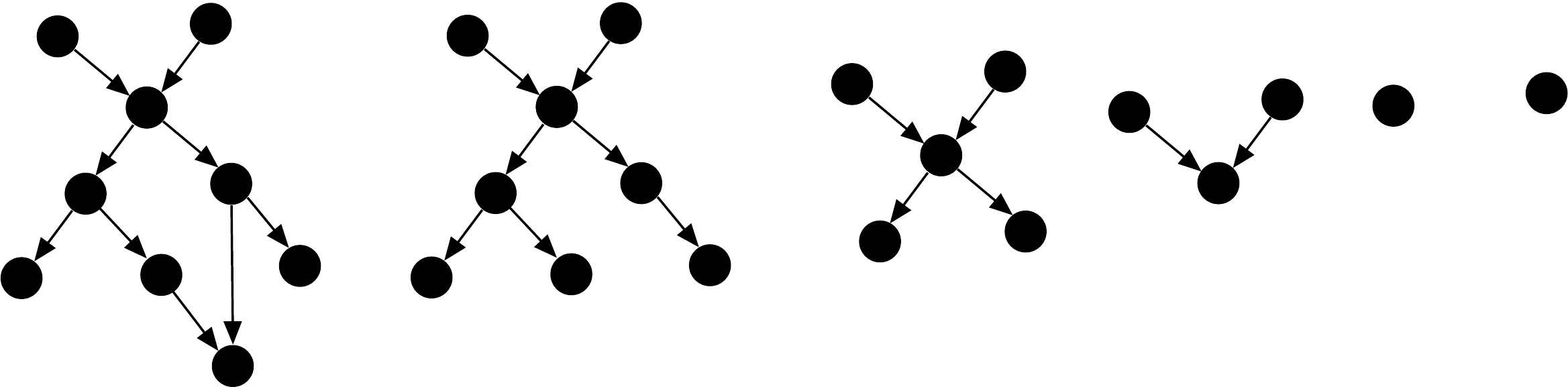}
\caption{The sequence of subgraphs ${\cal G}, {\cal G}_1,...,{\cal G}_4 $.}
\label{fig:SequenceBU}
\end{center}
\end{figure}

And the sequence of graphs ordered by inclusion ${\cal G}, \tilde{\cal G}_1,...,\tilde{\cal G}_4 $:
\begin{figure}[h]
\begin{center}
\includegraphics[width=8cm]{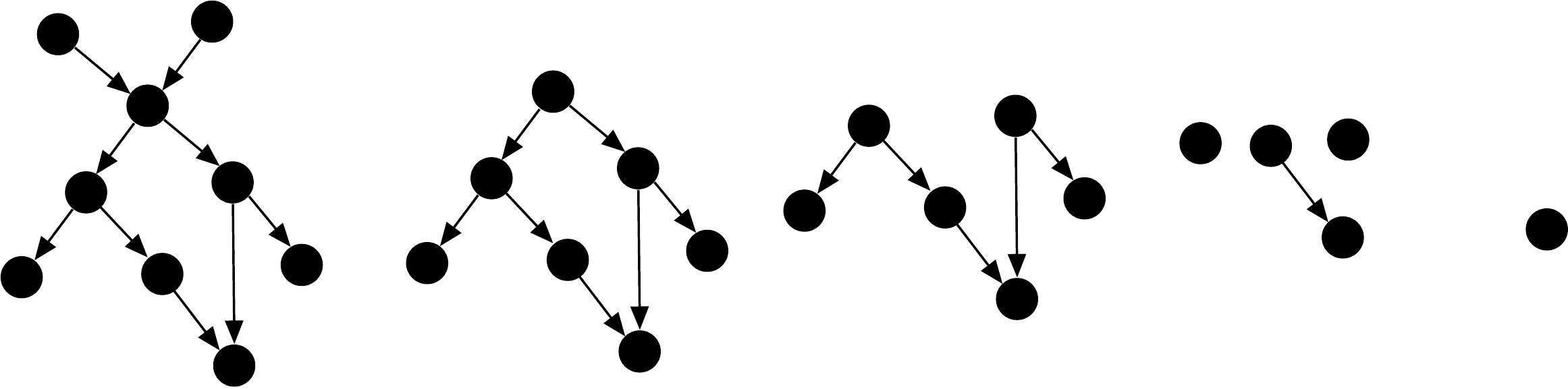}
\caption{The sequence of subgraphs ${\cal G}, \tilde{\cal G}_1,...,\tilde{\cal G}_4 $:}
\label{fig:LeafRemovalTD}
\end{center}
\end{figure}

\section{Detailed derivation of the Coordinates of Hierarchy}
\label{Sec:DetailedDerivation}

This section is devoted to the formalization of a quantitative descriptor of hierarchy. Such a descriptor is defined from  three hierarchy coordinates. These three coordinates are:  {\em Orderability} ($O$), {\em Feedforwardness} ($F$) and {\em Treeness} ($T$). Thus, any directed graph will define a point in such space, described as follows:
\[
\mathbf{u}({\cal G})\equiv (T({\cal G}), F({\cal G}), O({\cal G})).
\]
From the coordinates of $\mathbf{u}({\cal G})$ we can extract information about the structure of the net under the conceptual background of hierarchy presented in section \ref{Sec:Postulates} based on the concepts of {\em order}, {\em reversibility} and {\em pyramidal structure}. Through the values of these coordinates we must be able to identify and properly quantify deviations from such an ideal behavior. As we shall see, such three components naturally arise as long as we go in depth in our inquiry for a hierarchy estimator. The order $x=T, y=F,z=O$ derives from clarity issues related to the visualization. For the sake of clarity in the exposition, however, the order of the sections in which we present such measures will not follow such $T,F,O$ structure. Instead, we define the coordinates in such a way that we go from the simplest ($O$) to the most complex one ($T$).


We are now ready to define the coordinates of hierarchy.

\subsection{{\em Orderability}, $O$}
{\em This is the $z$ coordinate of $\mathbf{u}({\cal G})$}.

The {\em Orderability}, $O$, of the graph ${\cal G}$ is the fraction of nodes not belonging to any cycle -which are, by definition, non-orderable structures. In a more formal way, let ${\cal G}$ be a directed graph and ${\cal G}_{\cal C}$ be its condensed counterpart. The orderability of the graph $O({\cal G})$ is defined as:
\begin{equation}
O({\cal G})=\frac{|\{v_i\in V_{\cal C}\bigcap V\}|}{|V|}.
\label{eq:SCCF}
\end{equation}
In terms of the node-weighted condensed graph, we can rewrite the above expression as:
\[
O({\cal G})=\frac{|\{ v_i\in V_{\cal C}: \alpha_i=1\}|}{|V|}.
\]
 
Let us provide an example: In the graph depicted below (left), $|V=7|$, $|V_{\cal C}|=5$ (right). The grey circle depicts a condensed $SCC$. $|\{v_i\in V_{\cal C}\cap V\}|=4$, therefore, $O({\cal G})=4/7$:
\begin{figure}[h]
\begin{center}
\includegraphics[width=8cm]{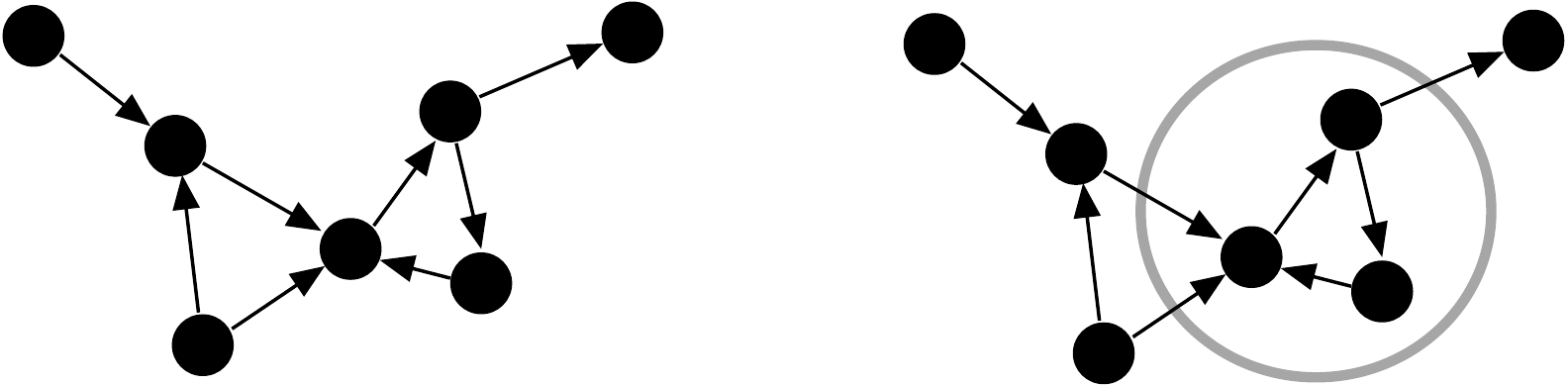}
\caption{Computing the orderability of the graph.}
\label{fig:Orderability}
\end{center}
\end{figure}

This is a raw indicator of the fraction of the net that can be ordered. A tree or, more generally, a feed-forward network will display $O({\cal G})=1$, in agreement to their orderable nature. In contrast, a network which is totally cyclical would display $O({\cal G})=0$.

\subsection{{\em Feedforwardness}, $F$}

{\em This is the $y$ coordinate of $\mathbf{u}({\cal G})$}.

Beyond the nodes than can be ordered, now we want to know the impact of the non-orderable regions of the graph over the potential causal paths described by it. In raw words, where, within the causal flow, we find the non-orderable regions. This is captured by the to-called {\em FeedForwardness}, $F({\cal G})$, a measure centered on the paths of the graph ${\cal G}_{\cal C}$ beginning in the set of maximal nodes, $M$ -see equations (\ref{MaximalSet}, \ref{minimalSet}) and (\ref{PiMmu}). Specifically, for every path going from   $M$ to $V_{\cal C}\setminus M$ \footnote{In the forthcoming lines, $M\equiv M({\cal G}_{\cal C})$, $\mu\equiv\mu({\cal G}_{\cal C})$, unless the contrary is indicated.}, we define a function, $F$, which evaluates the quotient between the number of nodes and the overall weight of the path. To cover all nodes of the graph, the numerical value of the coordinate will be averaged over the sequence of graphs obtained when applying a $LR_b$ algorithm. If  ${\cal G}_{\cal C}$ is finite both $M$ and $V\setminus M$ are finite and we can safely compute averages of these observables. 

To put the things in a more concrete way, let ${\cal G}$ be a graph and ${\cal G}_{\cal C}$ its condensed counterpart. If $\pi_k\in \Pi_{M\mu}({\cal G}_{\cal C})$, we define the function $F$ of $\pi_k$, $F(\pi_k)$, as:
\[
F(\pi_k)\equiv\frac{|v(\pi_k)|}{\sum_{v_i\in v(\pi_k)} \alpha_i}.
\] 
For example, if we take the graph studied above and we choose the highlighted path, to be named $\pi_j$:
\begin{figure}[h]
\begin{center}
\includegraphics[width=4.2cm]{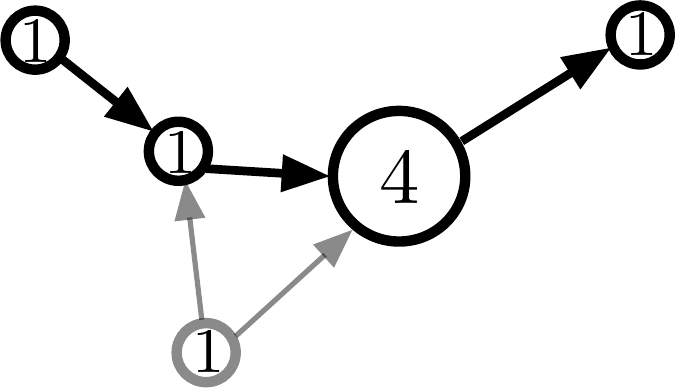}
\end{center}
\end{figure}

\noindent
we have that $F(\pi_j)=\frac{4}{7}$.

To compute the overall average we then build the function $g$:
\[
g({\cal G})=
\sum_{\pi_i\in \Pi_{M\mu}({{\cal G}_{\cal C}})}F(\pi_i),
\]
and we compute $g({\cal G}_1),...,g({\cal G}_{L({\cal G})-1})$ -see section \ref{Sec:Dissection} for the definition of ${\cal G}_i$'s:
\[
g({\cal G}_k)=
\sum_{\pi_i\in \Pi_{M\mu}({\cal G}_k)}F(\pi_i)
\]
We finally average over all ${\cal G}$'s and ${\cal G}_{\cal C}$, skipping ${\cal G}_{L({\cal G})}$ because, by its very definition, this graph has no paths:
\begin{equation}
F({\cal G})=\frac{g({\cal G}_{\cal C})+\sum_{k<L({\cal G}_{\cal C}) }g({\cal G}_k)}{{|\Pi_{M\mu}({\cal G}_{\cal C})|   + \sum_{k<L({\cal G}_{\cal C}) }| \Pi_{M\mu}({\cal G}_k)} |}.
\label{eq:FOP}
\end{equation}
And, for the sake of consistency, if the graph consists in a single node,
\[
(|V|=1)\Rightarrow (F({\cal G})=0).
\]
The interested reader can check that, with equation (\ref{eq:FOP}), we cover all paths from $M$ to $V\setminus M$.

Again, we will find $F({\cal G})=1$ in $DAG$s, as well as $F({\cal G})=0$ in networks consisting of a single strongly connected component. The combination of  $O({\cal G})$ and $F({\cal G})$ tells us how are cycles located within the net and the impact they have in the order of the structure. The combination of $O({\cal G})$ and $F({\cal G})$, thus, provides us interesting information on how the net is globally organized.

\subsection{\em Treeness, $T$}

{\em This is the $x$ coordinate of $\mathbf{u}({\cal G})$.}

This measure accounts for the {\em reversibility} and the {\em pyramidal structure}. The forthcoming coordinate is of information-theoretic nature and can only be computed over the condensed graph ${\cal G}_{\cal C}$, since it can only be defined over $DAG$s. The impact of cycles (non-orderable regions of the graph) is already evaluated by the two previous indicators.  This new coordinate largely relies on the computation of a special kind of entropies which, for the sake of brevity,  are not going to be derived in detail here. We refer the interested reader to \cite{Corominas-Murtra:2010,Corominas-Murtra:2011}. 

Intuitively, we first observe that deviations from reversibility can be properly captured measuring the uncertainty in reversing a given path, a measure called {\em topological reveresibility} \cite{Corominas-Murtra:2011}. This measure of uncertainty is provided by the statistical entropy over the set of paths present in a given graph. It turns out that deviations of the pyramidal structure can be also quantified through an information-theoretic measure, namely, from the difference between the amount of statistical entropy generated when we cross the graph according to the flow direction depicted by the arrows and the uncertainty in reversing the paths \cite{Corominas-Murtra:2011}. 
The spirit of the measure is to compare the creation of alternatives -new paths- in a top down exploration of the $DAG$ against the irreversibility of these paths. If the quantification of the uncertainty in reversing the paths  displays a lower value than the quantification of alternatives, we say that, qualitatively, the net shows a pyramidal structure. If the creation of new paths in a top down exploration of the graph is quantitatively equal to the uncertainty in reversing them, then, there is no argument to attribute to the net any pyramidal shape. Finally if the situation is opposite to the former one, the graph will display a funnel-like or inverted pyramid structure. Below we present the above mentioned entropies in more detail.

\subsubsection{Backwards entropy}

Let ${\cal G}$ be a directed graph and ${{\cal G}_{\cal C}}$ its condensed counterpart. Let $A({\cal G}_{\cal C})$ be the adjacency matrix of ${\cal G}_{\cal C}$. We first define the $|V_{\cal C}\setminus M|\times |V_{\cal C}\setminus M|$ matrix $\mathbf{B}({\cal G})$ in the following way:
\begin{equation}
(\forall v_i, v_j: A_{ij}=1)\;\;B({\cal G})_{ij}=\frac{A_{ij}({\cal G}_{\cal C})}{k_{in}(v_j)},
\label{B}
\end{equation}
and $(\forall v_i, v_j: A_{ij}=0)\;\;B({\cal G})_{ij}=0$.  From this definition, we obtain the
explicit dependency of the probability of crossing $v_j$ when we  start to reverse a path from $v_i$, $\mathbb{P}(v_j\leftarrow v_i)$. Thanks to the $DAG$-like nature of the condensed graph, we can compute such a probability directly from the powers of the adjacency matrix \cite{Corominas-Murtra:2010}, namely,
\begin{equation}
\mathbb{P}(v_j\leftarrow v_i)=\sum_{1\leq k\leq  L({\cal G}_{\cal C})}\left(\left[\mathbf{B}^T\right]^k({\cal
  G})\right)_{ij}.
\label{phi(B)}
\end{equation}
\begin{figure}
\begin{center}
\includegraphics[width=8cm]{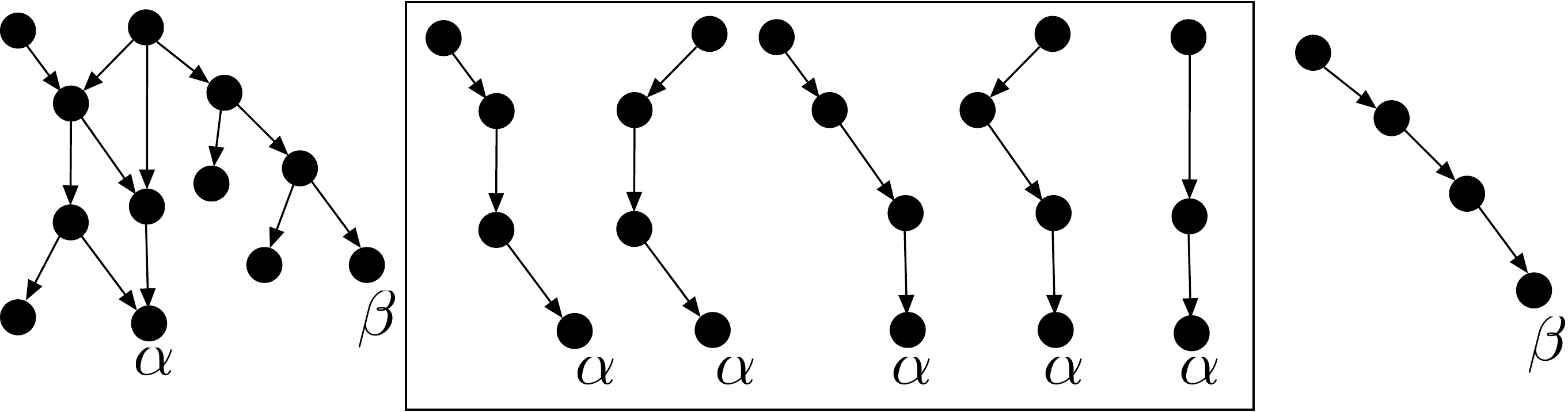}
\caption{We have a graph (left). Imagine that we know that some causal chain ended in node $\alpha$. We have $5$ different paths (center, inside the rectangle) to reach $\alpha$ from the top -the maximal nodes. So there is {\em uncertainty in reversing the path}. Finally (right), we observe that there is no uncertainty in reversing the causal path that ended in $\beta$, since there is only one path to go to $\beta$ from the maximal nodes. In this graph $H_b>0$ due to the uncertainty arisen when reversing the paths that ended in $\alpha$.}
\label{fig:BackwardsEntropy}
\end{center}
\end{figure}
Now we compute the average amount of uncertainty we have to face when {\em reversing} a path. Specifically, we have path starting at some node in $M$ and ending at node $v_j\in\mu$. We want to know the uncertainty of recovering this path if we go backwards, i.e., from the node $v_j\in \mu$ to a given node in $M$. This average uncertainty is provided by the following entropic functional \cite{Corominas-Murtra:2011}:
\begin{equation}
H_b({\cal  G}_{\cal C})=\frac{1}{|\mu|}\sum_{v_i\in      \mu}\sum_{v_k\in V_{\cal C}\setminus M}\mathbb{P}(v_i\leftarrow v_k)\cdot\log k_{in}(v_k).
\label{HGeneral}
\end{equation}

\subsubsection{Forward entropy}
\begin{figure}[h]
\begin{center}
\includegraphics[width=8cm]{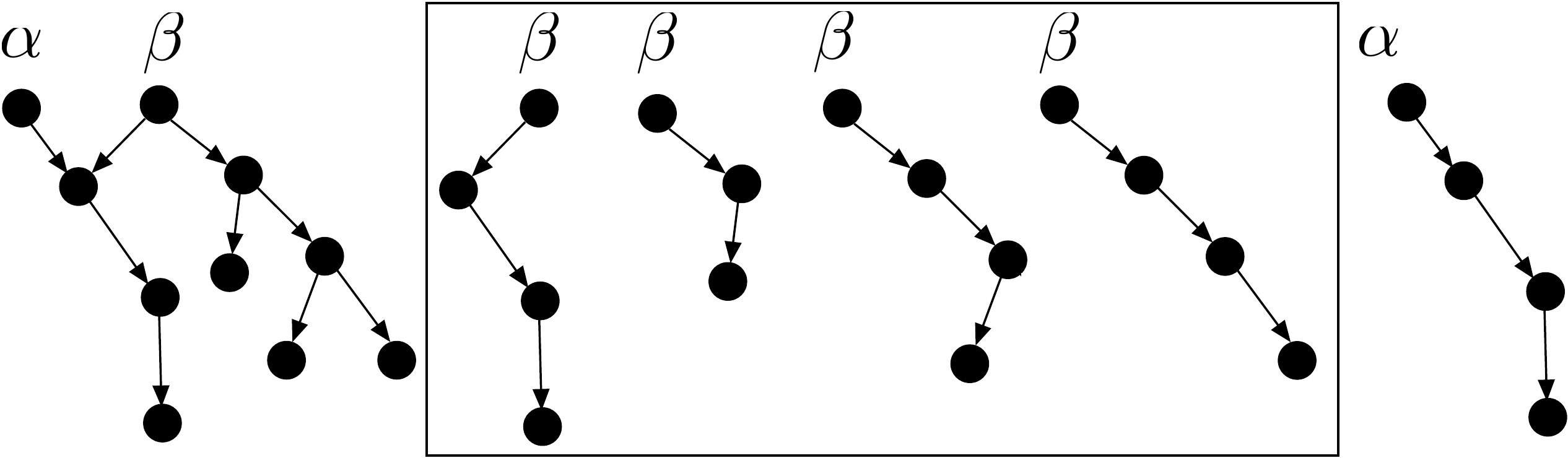}
\caption{We have a graph (left). Imagine that we know that some causal chain begun in node $\beta$. We have $4$ different paths (center, inside the rectangle) to reach a node belonging to the set of minimal nodes. So there is {\em uncertainty in following the path}. Finally (right), we observe that there is no uncertainty in following the causal path that begun in $\alpha$, since there is only one path to go from $\alpha$ to the minimal nodes. In this graph, $H_f>0$, due to the uncertainty arisen in following the paths starting in $\beta$.}
\label{fig:ForwardsEntropy}
\end{center}
\end{figure}
Now we compute the forward version of the above defined entropy. In this case, we need to compute the probability to cross node $v_k$ departing from $v_i\in M$ according to the causal flow -not reversing it, as above. The explicit expression of this probability is defined from matrix $\mathbf{B}'({\cal G})$:
\[
(\forall v_i, v_j: A_{ij}=1) \;\;B'({\cal G})_{ij}=\frac{A_{ij}({\cal G}_{\cal C})}{k_{out}(v_i)},
\]
and $(\forall v_i, v_j: A_{ij}=1)$,  $B'({\cal G})_{ij}=0$. Then, in analogy to what we found above -equation (\ref{phi(B)})-, we have that the probability to cross node $v_j$ if we started a path in $v_i$ is
\begin{equation}
\mathbb{P}(v_i\rightarrow v_j)=\sum_{1\leq k\leq  L({\cal G}_{\cal C})}\left(\left[\mathbf{B}'\right]^k({\cal G})\right)_{ij}.
\label{phi(B1)}
\end{equation}
The average of the amount of uncertainty emerging when we want to follow a path from which we know that starts in a given $v_i\in M$ and ends in a given node of $\mu$ will now be \cite{Corominas-Murtra:2011}:
\begin{equation}
H_f({\cal G}_{\cal C})=\frac{1}{|M|}\sum_{v_i\in M}\sum_{v_k\in V\setminus \mu}\mathbb{P}(v_i\rightarrow v_k)\cdot\log k_{out}(v_k),
\end{equation}
where $M$ is the set of maximal nodes.

\subsubsection{Treeness}
\begin{figure}[h]
\begin{center}
\includegraphics[width=8cm]{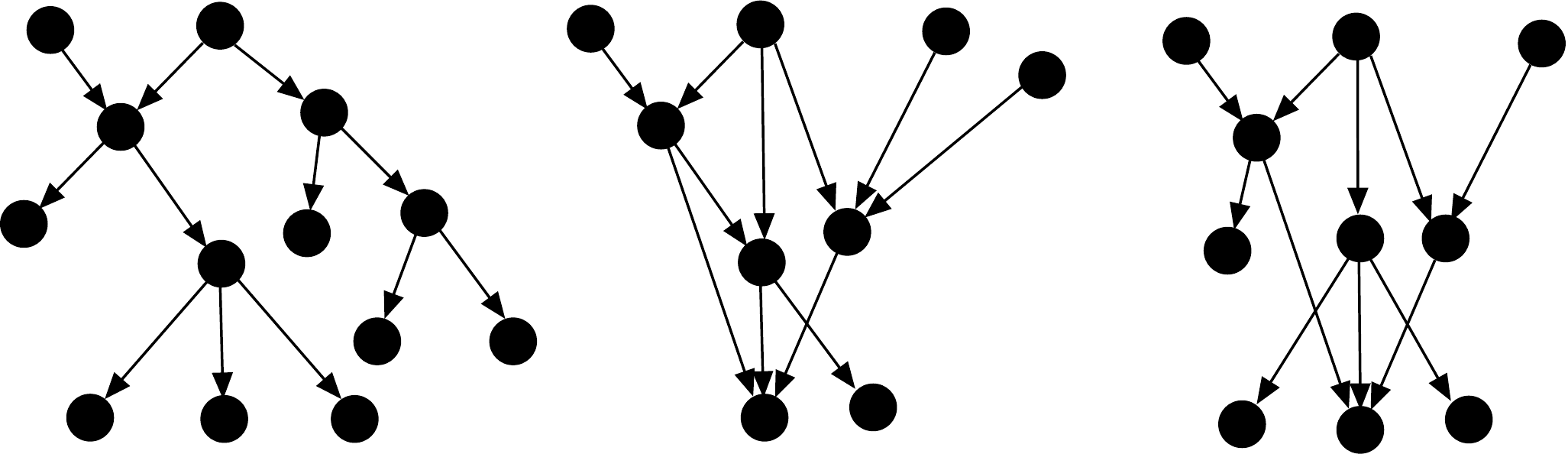}
\caption{Gaining intuition: Why do $H_b$ and $H_f$ grasp the hierarchical nature of graphs. The graph on the left displays $H_f-H_b>0$ and it is thereby hierarchical. On the contrary, the graph on the center displays $H_f-H_b<0$, being thus {\em anti}hierarchical. Finally, on the right we have a graph such that $H_f-H_b\approx 0$, thus it is neither hierarchical nor anitihierarchical.}
\label{fig:3DefGraphs}
\end{center}
\end{figure}
The presented coordinate compares the creation of new paths following the causal flow defined by the arrows against the uncertainty to reverse them. The creation of new paths is due to the existence of more than a single alternative to leave a given node. If we create new paths we are thus creating information -quantified by $H_f$. However, such information can be destroyed by the uncertainty in reversing the paths -evaluated by $H_b$. Intuitively, a graph having pyramidal structure will have the balance $H_f-H_b$ positive, whereas a graph having inverted pyramidal structure will display $H_f-H_b$ negative. A completely random $DAG$ displays $H_f-H_b\approx 0$ \cite{Corominas-Murtra:2011}. It turns out that, properly manipulated, such information measures are the perfect indicators of the pyramidal properties of the graph, a key ingredient of our hierarchy coordinates.

In order to generate a normalized estimator (between $-1$ and $1$) accounting for the balance between 
$H_f({\cal G}_{\cal C})$ and $H_b({\cal G}_{\cal C})$ we define $f({\cal G})$  as follows:
\begin{equation}
f({\cal G})\equiv \frac{H_f({\cal G}_{\cal C})-H_b({\cal G}_{\cal C})}{\max\{ H_f({\cal G}_{\cal C}),H_b({\cal G}_{\cal C})\}}.
\label{eq:f}
\end{equation}
The {\em treeness coordinate  of a $DAG$}, to be indicated as $T({\cal G})$, will be the average among the $L({\cal G}_{\cal C})-1$ subgraphs ${\cal G}_1,...,{\cal G}_k,...,{\cal G}_{L({\cal G}_{\cal C})-1}$, the $L({\cal G}_{\cal C})-1$ subgraphs $\tilde{\cal G}_1,...,\tilde{\cal G}_k,...,\tilde{\cal G}_{L({\cal G}_{\cal C})-1}$ and ${\cal G}$ itself, i.e., the set ${\cal W}({\cal G})$ defined in equation (\ref{eq:DefW}) \footnote{Note that i) we ruled out the contributions of ${\cal G}_{L({\cal G}_{\cal C})}$ and $\tilde{\cal G}_{L({\cal G}_{\cal C})}$ because they contain no links by definition, and, therefore, would impact to the computation of the hierarchy without justification, and ii) consistently, we compute the average between $2L({\cal G}_{\cal C})-1$ objects, the sie of the set ${\cal W}({\cal G})$ defined in equation (\ref{eq:DefW}).}, i.e.:
\begin{eqnarray}
T({\cal G})&=&\frac{1}{2L({\cal G}_{\cal C})-1}\left(f({\cal G})+\sum_{i<  L({\cal G}_{\cal C})}f({\cal  G}_i)+f(\tilde{\cal G}_i)\right)\nonumber\\
&=&\frac{1}{|{\cal W}_{\cal G}|}\sum_{{\cal G}_i\in {\cal W}({\cal G})} f({\cal  G}_i)\nonumber\\
&=&\langle f\rangle_{{\cal W}_{\cal G}},
\label{hierarchy}
\end{eqnarray}
where ${\cal W}_{\cal G}$ is the set of subgraphs of ${\cal G}_{\cal C}$ obtained through the application of a lead removal algorithm (either bottom up or top down) and which contain at least one link, as defined in equation (\ref{eq:DefW}).
\begin{figure}[h]
\begin{center}

\includegraphics[width=8cm]{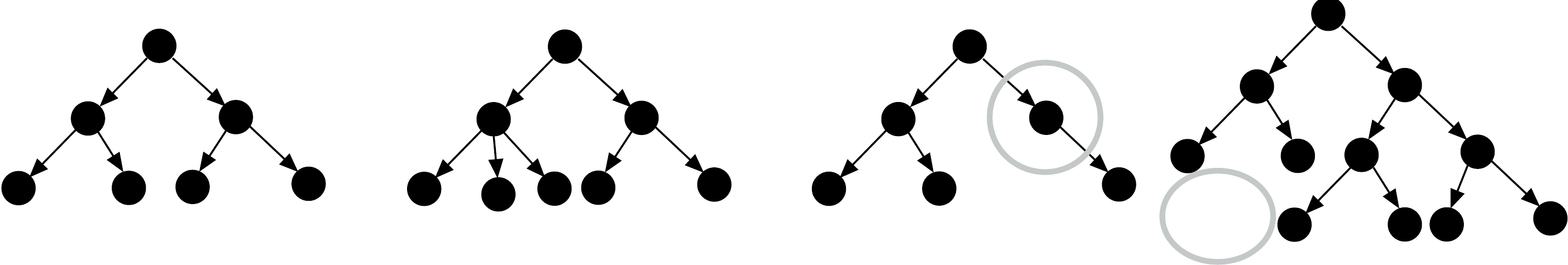}
\caption{The need for detailed layer analysis. All four graphs display $H_f>0$ and $H_b=0$. However, the two graphs on the right contains a violation on the hierarchy assumptions, namely, the need for {\em pyramidal} structure at all levels. The dissection allows to identify and penalize such deviations. }
\label{fig:HierarchyViolations}
\end{center}
\end{figure}

For the sake of consistency we explicitly define the behavior of some limit, potentially problematic cases.
When obtaining the sequence of ${\cal G}$'s and $\tilde{\cal G}$'s the graph can break into more than a single connected component. 
Let $\Gamma({\cal G}_i)=\{\gamma_1,...,\gamma_k\}$  the set of components of our graph ${\cal  G}_i$, let $V^i(\gamma_j)$ be the set of nodes of the $j$-th component of ${\cal G}_i$ and let $\tilde{V}^i(\gamma_j)$ the set of nodes of the  $j$-th component of $\tilde{\cal G}_i$. The parameter $f({\cal G}_i)$ is evaluated averaging the 
individual  contributions  of the  different  connected components  of
${\cal G}_i$  or $\tilde{\cal G}_i$  according to the number  of nodes
they have against $|V_i|$ or $|\tilde{V}_i|$, the set of all nodes of ${\cal G}_i$ and $\tilde{\cal G}_i$, respectively, leading to:
\[
f({\cal G}_i)\equiv\frac{1}{|V_i|} \sum_{\Gamma({\cal G}_i)}|V^i(\gamma_k)|f(\gamma_k)\;\;{\rm and}
\] 
\[
f(\tilde{\cal G}_i)\equiv\frac{1}{|\tilde{V_i}|} \sum_{\Gamma(\tilde{\cal G}_i)}|\tilde{V}^i(\gamma_k)|f(\gamma_k).
\] 

We  impose,  for  both  mathematical  and
conceptual consistency, that:
\[
(\max\{H_b({\cal G}_{\cal C}),H_f({\cal G}_{\cal C})\}=0)\Rightarrow (T({\cal G})\equiv 0).
\]
Furthermore,   if
$E_{\cal C}=\varnothing$, (i.e., the case where  the graph consists of a single
node):
\[
T({\cal G})\equiv 0.
\]

\begin{figure*}
\begin{center}
\includegraphics[width=9cm]{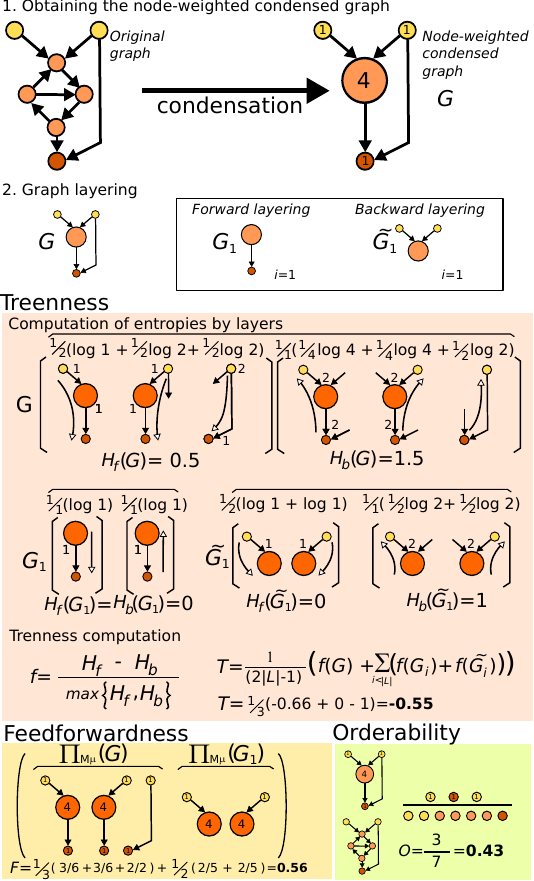}
\caption{The definition of the hierarchy-based morphospace $\Omega$. Visual representation of the computation of the three axes of the morphospace: treeness, feedforwardness and orderability ($TFO$ coordinates, $\mathbf{u}$).}
\label{fig:1}
\end{center}
\end{figure*}
\begin{figure*}
\begin{center}
\includegraphics[width=11cm]{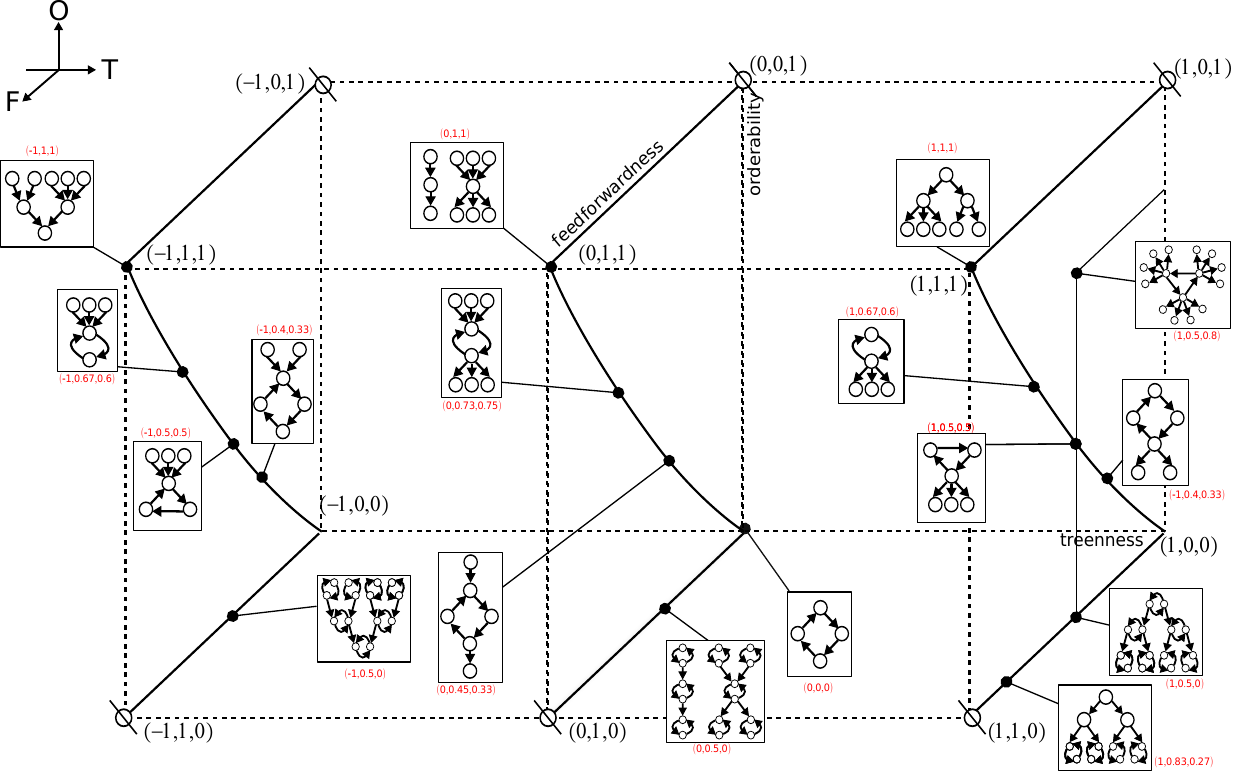}
\caption{Illustration of the kind of graphs living in the morphospace $\Omega$. We located different kinds of toy graphs in their corresponding $TFO$ coordinates, $\mathbf{u}$, to grasp the kind of structures one might expect to observe in the different regions of the morphospace. Numbers in red indicate $\mathbf{u}$ for every graph.}
\label{fig:toys}
\end{center}
\end{figure*}

%

\subsection{The coordinates of hierarchy for directed graphs}
These three coordinates enabled us to define a morphospace, $\Omega$, in which every net will be represented by a point 
$\mathbf{u}({\cal G})= (T({\cal G}), F({\cal G}), O({\cal G}))$ in this space.
The coordinates defining such space are thought of to grasp the essentials of {\em hierarchy}, taking as the canonical hierarchical structure the one satisfying the postulates presented in section \ref{Sec:Postulates}. Additionally, as we have seen, comparison between coordinates also provides us valuable information concerning the whole structural organization of the net. For example, the relation between $O$ and $F$ enables us to estimate how many cyclic regions are in the graph and at what point of the causal flow they are located. In addition, they provide an estimator of relevance of coordinate $T$, since low values of $O,F$ indicate us that the net is mainly cyclical and that the feed-forward-like analysis provided by $T$ is less relevant than in the case where $O, F$ are close to $1$. To test the consistency of this coordinate system, we will rigorously check the behavior in several extreme and paradigmatic cases:

\noindent
{\bf Lemma}:{\em {\cal G} displays $T({\cal G})=F({\cal G})=O({\cal G})=1$ if and only if ${\cal G}$ is a  directed tree  in which all the leafs have the same length, as defined in section \ref{Sec:DirectedGraphs}.}

\noindent{\bf Proof}$\\$
$(\Rightarrow)$ By definition, this graph is a $DAG$, therefore,  $F({\cal G})=O({\cal G})=1$. Furthermore, all its nodes but the one in $M$ have $k_{in}=1$, which means that
\begin{equation}
(\forall {\cal G}'\in \{{\cal G}_1,,{\cal G}_{L({\cal G}_{\cal C})-1}\}\bigcup\{\tilde{\cal G}_1,...,\tilde{\cal G}_{L({\cal G}_{\cal C})-1}\} )(H_b({\cal G}')=0).
\label{eq:H_b=0}
\end{equation}
And, since all but the set of minimal nodes will display $k_{out}\geq 2$, then:
\begin{equation}
(\forall {\cal G}'\in \{{\cal G}_1,,{\cal G}_{L({\cal G}_{\cal C})-1}\}\bigcup\{\tilde{\cal G}_1,...,\tilde{\cal G}_{L({\cal G}_{\cal C})-1}\} )(H_f({\cal G}')>0),
\label{eq:H_b>0}
\end{equation}
and, thus,
\begin{widetext} 
\[
(\forall {\cal G}'\in \{{\cal G}_1,,{\cal G}_{L({\cal G}_{\cal C})-1}\}\bigcup\{\tilde{\cal G}_1,...,\tilde{\cal G}_{L({\cal G}_{\cal C})-1}\} )(\max\{H_f({\cal G}'),H_b({\cal G}')=H_f({\cal G}')).
\]
\end{widetext}
The above claims are also true for the whole graph ${\cal G}$ Therefore, all the $2L({\cal G}_{\cal C})-1$ quotients of the type shown in equation (\ref{eq:f}) involved in the computation of $T({\cal G})$ will have a value value equal to $1$, which leads to $T({\cal G})=1$.

\noindent
$(\Leftarrow)$ If $F({\cal G})=O({\cal G})=1$, then ${\cal G}$ is a $DAG$. To see that $T({\cal G})=1$, it is enough to realize that the graph must satisfy equations (\ref{eq:H_b=0}, \ref{eq:H_b>0}) and that this can only happen if ${\cal G}$ is a directed tree  in which all the leafs have the same length. We observe that, if the laves have not the same length, then a $0$ will emerge in some computation, lowering the value to $T({\cal G})<1$.

\noindent{\bf Corollary}: {\em ${\cal G}$ displays $T({\cal G})=-1,F({\cal G})=O({\cal G})=1$ if and only if ${\cal G}$ is a  directed tree  in which all the leafs have the same length, as defined in section \ref{Sec:DirectedGraphs} but with changing the direction of all arrows.}

This latter case would belong to the perfect {\em antihierachical} system. A graph consisting in a big cycle will be the paradigmatic example of {\em non-hierarchical} system, showing thus $T({\cal G})=F({\cal G})=O({\cal G})=0$. Finally, a graph consisting of a chain of nodes, or a $DAG$ in which all the layers have the same size and all nodes the same connectivity will be the paradigmatic example of {\em non-hierarchical} but {\em ordered system}, showing $T({\cal G})=0, F({\cal G})=O({\cal G})=1$. An illustration of the gallery of conformation on the region of possible networks in the $\Omega$ space is provided in figure (\ref{fig:toys}).  Notice that some regions of the morphospace cannot be occupied by the very definition of the measures, while others represent strange configurations.  It is worth to note that it is mathematically possible to apply this formalism to networks containing more than one connected component. However, meaningful information can only be obtained from the study of a single component since it represents a unique causal structure.

In figure (\ref{fig:1}) we detail the computation of the hierarchy coordinates of a given graph. From the original  directed graph, the process of strongly connected component ($SCC$) detection and condensation give rise a node-weighted condensed graph by from the original graph. From this graph, the three coordinates are computed.

\subsection{Analytical estimates for random graphs}

We will now provide estimations for the hierarchy coordinates in random directed graphs. A word of caution is needed. Exact calculations of the above presented values are hard and far away from the scope of this work. We therefore will present the analytic results starting from some simplifying assumptions to obtain indications on what we should expect in the case of large, sparse random networks. We will focus on $T$ and $O$, since the active presence of paths and cycles in the computation of $F$ makes any rough estimation extraordinarily complex and of no practical use.

\subsubsection{$T$ in random directed graphs}
\label{sec:RandomAnalytics}

We are going to show that the expected value $\langle T\rangle$ over an ensemble of random directed graphs is $0$. The reasoning takes advantage of the internal symmetries of the ensemble and uses several critical assumptions which will be clearly highlighted. A totally rigorous derivation of this result would require a deep exploration of such assumptions, something that goes far from the main scope of the presented work.

Let us suppose that we have an undirected graph ${\cal G}$ having adjacency matrix $A({\cal G})$. Now, {\em every undirected link} is transformed into a directed one and the direction of the arrow is defined at random with probability $p=1/2$. Let us define $G_d({\cal G})$ as the ensemble of all $M$ possible  directed graphs we can build from ${\cal G}$ by imposing a direction at random over the links present in such  graph: 
\[
G_d({\cal G})=\{{\cal G}_1^d, . . .,{\cal G}_M^d\}.
\]
Let now $A_1^d,. . .,A_M^d$ be their respective adjacency matrices. In addition, let $\left({\cal G}_i^d\right)^T$ be the graph described by the transpose of the adjacency matrix of ${\cal G}_i^d$, $\left(A_i^d\right)^T$. We observe that, since directions have been assigned at random:
\[
({\cal G}_i^d\in G_d({\cal G}))\Rightarrow\left( \left({\cal G}_i^d\right)^T\in G_d({\cal G}\right).
\]
For any graph we can build, a graph with all the directions of the links switched can also be built up.We can therefore induce a partition in $G_d({\cal G})$, $\tilde{G}_d({\cal G})$, made of pairs of graphs of $G_d({\cal G})$, namely:
\begin{equation}
\tilde{G}_d({\cal G})=\{\{{\cal G}_1^d,\left({\cal G}_1^d\right)^T\}, . . .,\{{\cal G}_m^d,\left({\cal G}_m^d\right)^T\}\},
\label{eq:Partition}
\end{equation}
and, since this is a partition of $G_d({\cal G})$, we highlight that
\[
(\forall x,y\in \tilde{G}_d({\cal G}))\left(x\bigcap y =0\right) ;\;\;\;\bigcup \tilde{G}_d({\cal G})={G}_d({\cal G});\;\;\; m=\frac{M}{2}.
\]
These properties will be useful in the forthcoming derivations.
We observe that the partition of the ensemble in pairs of graphs whose adjacency matrices are mutually transposed will also be possible after the condensation operation, therefore,
if $G_{\cal C}({\cal G})$ is the ensemble of all possible condensed graphs out of all directed graphs composing $G_d({\cal G})$:
\[
G_{\cal C}({\cal G})=\{{\cal G}_1^{\cal C}, . . .,{\cal G}_{M'}^{\cal C}\},
\]
(notice that $M'< M$ in most cases), then, 
\begin{equation}
({\cal G}_i^{\cal C}\in G_{\cal C}({\cal G}))\Rightarrow\left( \left({\cal G}_i^{\cal C}\right)^T\in G_{\cal C}({\cal G}\right),
\label{eq:TransponseDAG}
\end{equation}
also holds. This symmetry within the ensemble $ G_{\cal C}({\cal G})$ is found as long as $p=1/2$, i.e., directions of links are totally imposed at random. We keep this in mind and we proceed to compute the expected value of the backwards and forward entropy over the ensemble, $\langle H_f\rangle,\langle H_b\rangle$, and we see that:
\begin{eqnarray}
\langle H_f\rangle&=&\frac{1}{|G_{\cal C}({\cal G})|}\left[\sum_{{\cal G}_{\ell}^{\cal C}\in G_{\cal C}({\cal G}))}\right.\nonumber\\
&&\left.\frac{1}{|M_{\ell}|}\sum_{v_i\in M_{\ell}}\sum_{v_k\in V\setminus \mu_{\ell}}\mathbb{P}(v_i\rightarrow v_k)\cdot\log k_{out}(v_k)\right]\nonumber\\
&=&\frac{1}{|G_{\cal C}({\cal G})|}\left[\sum_{{\cal G}_{\ell}^{\cal C}\in G_{\cal C}({\cal G}))}\right.\nonumber\\
&&\left.\frac{1}{|M_{\ell}|}\sum_{v_i\in \mu_{\ell}}\sum_{v_k\in V\setminus M_{\ell}}\mathbb{P}(v_i\leftarrow v_k)\cdot\log k_{in}(v_k)\right]\nonumber\\
&=&\langle H_b\rangle\nonumber,
\end{eqnarray}
where $M_{\ell}$ and $\mu_{\ell}$ stand for the maximal and minimal sets of ${\cal G}_{\ell}^{\cal C}$. Notice that the crucial step is the second equality, where we use the symmetry of $G_{\cal C}({\cal G})$ described in equation (\ref{eq:TransponseDAG}). We therefore have that
\[
\langle H_f\rangle=\langle H_b\rangle.
\]
This tells us that $\langle H_f\rangle-\langle H_b\rangle=0$. However, we cannot jump directly to conclude that $\langle H_f-H_b\rangle=0$. 
Using the same reasoning we used above, we compute $\langle H_f-H_b\rangle$:
\begin{eqnarray}
\langle H_f-H_b\rangle&=&\frac{1}{|G_{\cal C}({\cal G})|}\sum_{{\cal G}_i^{\cal C}\in G_{\cal C}({\cal G})}H_f({\cal G}_i^{\cal C})-H_b({\cal G}_i^{\cal C})\nonumber\\
&=&\frac{1}{|G_{\cal C}({\cal G})|}\sum_{\{{\cal G}_i^{\cal C},({\cal G}_i^{\cal C})^T\}\in \tilde{G}_{\cal C}({\cal G})}H_f({\cal G}_i^{\cal C})-H_b(({\cal G}_i^{\cal C})^T)\nonumber\\
&=&0.\nonumber
\end{eqnarray}
Again, in the second step, we rearranged the terms of the sum  by means of the partition $\tilde{G}_{\cal G}({\cal G})$ -see equation (\ref{eq:Partition})- induced over the ensemble $G_{\cal C}({\cal G})$, thereby resetting all terms of the sum to zero. Thus, we can conclude that:
\begin{eqnarray}
\langle f\rangle&=&\left\langle \frac{ H_f- H_b}{\max\{  H_f, H_b\}}\right\rangle\nonumber\\
&=&0.\nonumber
\end{eqnarray}
And, since it is true {\em in general}, we have proven that given a graph ${\cal G}$, if we build an ensemble of random directed graphs $G_d({\cal G})$ using the procedure described at the beginning of the section, then:
\begin{equation}
\langle T\rangle=0.
\label{eq:PredictioT}
\end{equation}
A couple of remarks are in force. $i)$ The first one concerns the assumption that the ensemble $G_{\cal C}({\cal G})$ is {\em well behaved}, which means that the averages collapse to the most probable value of a given observable. As we shall see, this assumption holds in general in ensembles of random graphs, and it seems reasonable to assume that this is independent of the degree distribution, as long as the graph is obtained using the random  procedure presented above. $ii)$ The second remark concerns cycles containing $2$ nodes which, by construction, are avoided in the above developments. However, we observe that these bidirectional links already introduce a symmetry within the adjacency matrix under the transpose operation and, consequently, their presence would have no impact in equation (\ref{eq:TransponseDAG}) and thus the reasoning still holds. We remark that this is what is observed in random graphs, independently of their degree distributions, as we see in figure (15) and in the forthcoming sections.
\begin{figure}
\includegraphics[width=8.2cm]{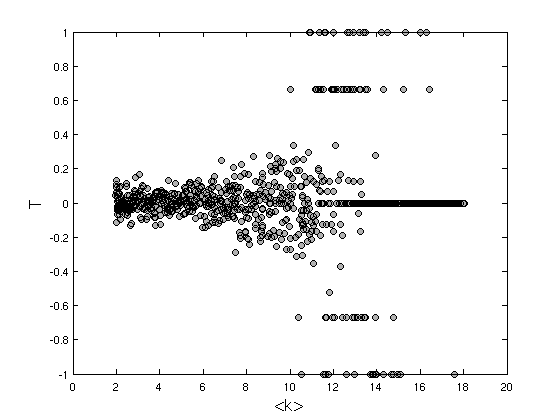}
\caption{Evolution of the coordinate $T$ in an ensemble of directed random graphs of size $n=1000$ in terms of the average degree $\langle k\rangle$. As predicted by equation (\ref{eq:PredictioT}), values accumulate around $T=0$. As the connectivity increases, the variance increases as well. This is due to the process of condensation, that reduces drastically the size of the graph, since most nodes belong -as we shall see in the next section- to a single cycle. Therefore, the resulting condensed graph is very small and even being maximals directly connected to minimals, thereby revealing, in some cases, extreme values of $T$, like $-1$ or $1$.}
\label{fig:PlotTTheoretical}
\end{figure}

\subsubsection{$O$ in random directed graphs}

Now we will obtain an estimate of the evolution of the $O$ coordinate. 
To this end we will use standard theory of generating functions applied to the emergence of giant components within a random graph. We will not develop the reasoning since this is far from the scope of the work and they it is clearly developed in \cite{Dorogovtsev:2001}, \cite{Newman:2001}  or \cite{Newman:2010}. We recommend the interested reader to go to this literature and references therein.

Let us work with an ensemble of directed random graphs, ${\cal G}(V,E)$. The undirected average degree will be thus:
\[
\langle k\rangle =\frac{2|E|}{|V|}
\]
and the average $in$ and $out$ degree will be
\[
\langle k_{in}\rangle =\frac{|E|}{|V|}=\langle k_{out}\rangle 
\]
i.e., since directions are spread at random, we assume that $\langle k_{in}\rangle=\langle k_{out}\rangle\approx\frac{\langle k\rangle}{2}$.
It is well known that, in general, after a certain threshold of $\langle k\rangle$ the {\em Giant Connected Component} ($GCC$) emerges, namely, a connected component of ${\cal G}$ containing a finite fraction,  ${S}$, of the nodes of the graph  \cite{Newman:2001}. For directed graphs, in addition, we observe further the emergence of the {\em Giant Strongly Connected Component}, ($GSCC$) namely, a single $SCC$  containing a finite fraction  $\mathbf{S}$ of the nodes of the graph \cite{Dorogovtsev:2001}. The $GSCC$ will represent the largest {\em cyclic region} of the graph. Consequently, if we assume that small $SCC$s that can be also present within the graph represent a negligible fraction of it when compared to the size of the $GSCC$, one can approach the following:
\begin{equation}
O\approx 1-\frac{\mathbf{S}}{S}.
	\label{eq:PredictioO}
\end{equation}
The $1/S$ term acts as a normalization factor over the connected fraction of the graph, since our hierarchy computations only make sense over fully connected structures. Notice that the probability to have $2$ $GCC$ is vanishingly small \cite{Newman:2010} and thus, with high probability
\[
GSCC\subset GCC,
\]
therefore, the normalization of the relative size of the $GSCC$ with the size of the $GCC$ provides us an estimation about the actual impact of cycles over the connected structure of the graph.
We then use generating function methodology to obtain an estimate both $S$ and $\mathbf{S}$. Let $f(r)$ be the generating function of the degree distribution, $p(k)$, of the graph ${\cal G}$ having average degree $\langle k\rangle$:
\[
f( r)=\sum_{k}p(k)r^k.
\]
Then, following \cite{Newman:2001}, we have that
\begin{equation}
S=1-f(r_c),
\label{eq:Sgiant}
\end{equation}
being $r_c$ the smallest positive solution of
\[
r_c=\left.\frac{1}{\langle k\rangle}\frac{\partial}{\partial r}f( r)\right|_{r=r_c}.
\]
To obtain $\mathbf{S}$ of the random directed graph, one proceeds in an analogous way:
Let now $g(x,y)$ be the generating function of the joint $in$-degree and $out$-degree distribution, $p(k_{in},k_{out})$ of a given random graph ${\cal G}$:
\[
g(x,y)=\sum_{k_{in}, k_{out}}p(k_{in}, k_{out})x^{k_{in}}y^{k_{out}}.
\]
Following \cite{Dorogovtsev:2001}, we know that, if there is no correlation between $in$ and $out$-degrees $p(k_{in},k_{out})\approx p(k_{in})p(k_{out})$, we can estimate the size of the $GSCC$ as:
\begin{equation}
\mathbf{S}=(1-g(x_c,1))(1-g(1,y_c)),
\label{eq:SGCC}
\end{equation}
where $x_c$ is smallest positive root of the following self-consistent equation:
\[
x_c=\left.\frac{1}{\langle k_{in}\rangle}\frac{\partial}{\partial y}g(x,y)\right|_{x=x_c,y=1},
\]
and, identically, $y_c$ is smallest positive root of the following self-consistent equation:
\[
y_c=\left.\frac{1}{\langle k_{out}\rangle}\frac{\partial}{\partial x}g(x,y)\right|_{x=1,y=y_c}.
\]
We emphasize that this independence condition among $k_{in}$ and $k_{out}$ is a very strong one, consequently, results must be seen as a rough estimation of the qualitative behavior of the $GSCC$. 

To obtain concrete results, we now turn to an ensemble of directed $ER$ graphs with uncorrelated $in$ and $out$ degrees. In this ensemble , $|E|$ directed links are spread at random among pairs of the $|V|$ existing nodes. The $in$ and $out$ average degrees will be $\langle k_{in}\rangle=\langle k_{out}\rangle\approx\frac{\langle k\rangle}{2}$, and the degree distribution of the $in$ and $out$ binomial distribution around the above mentioned averages. Under this framework, one has that:
\[
g(x,1)=e^{\frac{\langle k\rangle}{2}(x-1)};\;\;g(1,y)=e^{\frac{\langle k\rangle}{2}(y-1)};
\]
and
\[
f ( r )=e^{{\langle k\rangle}(r-1)}.
\]
Thereby obtaining an interesting simplification of our problem, namely,
\[
\frac{1}{\langle k_{in}\rangle}\frac{\partial}{\partial y}g(x,y)=g(x,y)=\frac{1}{\langle k_{out}\rangle}\frac{\partial}{\partial x}g(x,y).
\]
Thus, the above critical values read:
\[
x_c=e^{\frac{\langle k\rangle}{2}(x_c-1)}=y_c,
\]
and, from equation (\ref{eq:PredictioO}),
\begin{equation}
O=1-\frac{(1-x_c)^2}{1-r_c},
\label{eq:Oestimate}
\end{equation}
since, from equation (\ref{eq:Sgiant}) $S=1-r_c$.

In figure (16) we plot estimates of equation (\ref{eq:Oestimate}) and real values from an ensemble of Erdos-Renyi graph and we see that, despite the strong assumptions made, the behavior of $O$ can be clearly predicted from the mentioned equation, providing a good insight to the behavior of such coordinate.
\begin{figure}
\includegraphics[width=8.2cm]{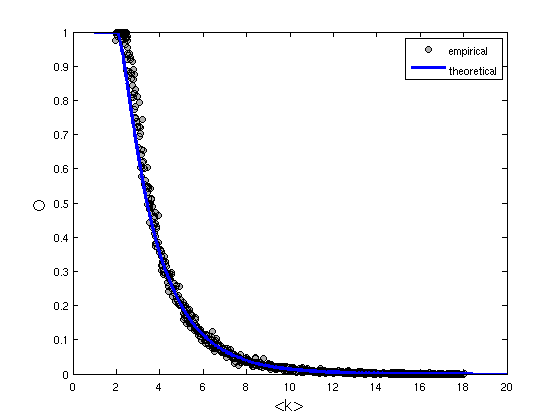}
\caption{Evolution of the coordinate $O$ in an ensemble of directed random graphs of $|V|=1,000$ in terms of the average degree $\langle k\rangle$. Similar behaviour was obtained for ensembles of 500 nodes. The blue line contains the theoretical prediction provided by equation (\ref{eq:Oestimate}) and dots represent actual graphs of the ensemble. As we can see the fitting between real data and the theoretical prediction is, in spite the assumptions made, very good.}
\label{fig:PlotOTheoretical}
\end{figure}

\subsubsection{$F$ in random directed graphs}
In order to get insight about the behaviour of this coordinate we provide a numerical computation of F for an ensemble of increasing directed ER networks. Figure (17) shows the dramatic impact of $\langle k \rangle$ in $F$. In this case, due to $F$ intrinsically depends on the diameter of the graph in the computation of the number of pathways from minimal  to maximals in the condensed graph.  It is not ignored by the authors that the drastic reduction of $F$ by increasing $\langle k\rangle$ is tied to the impact of $SCC$ size affecting to the length of pathways in the condensed graph. However, we consider that to find this connection go beyond the aim of a work that pursue the presentation of a formalisation and characterization of the concept of hierarchy in the framework of a morphospace. Further work in this direction will contribute to the analytical comprehension of the impact of  $\langle k \rangle$ in $F$.


All the theoretical work is thus finished. From now on, we will apply such machinery to the analysis of both model and real networks. 
\begin{figure}
\includegraphics[width=8.2cm]{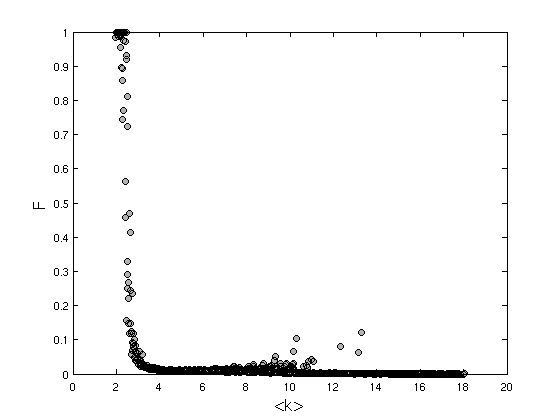}
\caption{Evolution of the coordinate $F$ in an ensemble of directed random graphs of $|V|=1,000$ in terms of the average degree $\langle k\rangle$. Similar behaviour was obtained for ensembles of 500 nodes.Note the slight data dispersion at high $\langle k \rangle$ values. This region correspond to connectivities where the resulting condensed graph is very little. In such  circumstance, variations in the position of the SCC in node weighted condensed graph has a considerable impact of $F$ computation.}
\label{fig:PlotFempirical}
\end{figure}

\section{Analysis of Networks}

In this section we systematically study the location of both real and model networks in the morphospace defined by the three coordinates $T({\cal G}), F({\cal G}), O({\cal G})$. We begin  by studying classical models of random graphs, namely, the Erd\"os-R\'enyi (ER) \cite{Erdos:1960}, the Barab\'asi-Albert preferential attachment (BA) \cite{Barabasi:1999} and the Callaway-Hopcroft-Kleinberg-Newman-Strogatz   uniform attachment (Callaway) \cite{Callaway:2001}.  Then, we evaluate the hierarchy coordinates of $125$ real networks belonging to different systems. The analysis of real nets is completed by confronting the obtained results with the exploration of the hierarchy coordinates of their corresponding randomized ensembles. We use two randomization methods which are exposed in detail. Finally we present an in silico experiment of evolution of networks inside the morphospace. Such analysis enables us to explore the accessibility of the space of possible configurations and sheds light on what is likely to observe in real systems.

\subsection{Model Networks} 
\label{Model networks}

Analysis of model networks obtained using the directed versions of ER, BA and Callaway models show similar results. In all cases, model networks having high connectivities aggregate around the region defined by the rectangle $T=(-1,1)$, $F\approx (0,0.5)$ $O\approx0$. The high variance on the value of $T$ can be explained by the fact that
${\cal G}_{\cal C}$ is very small, since high connectivity produces a drastic process of condensation. It is easy to see that a network with a handful of nodes is more likely to give extreme values than large networks.  Departing from the value $O\approx 0$ displayed at high connectivities, as connectivity decreases, $O$ starts to increase but nets still aggregate around $T\approx0$ and $F\approx0$. For low connectivities, $T$ values tend to be closer to $T=0$ but both $F$ and $O$ change their behavior, increasing their values to reach the region around the point $T=0, F=1, O=1$. This latter situation can be explained by the low presence -or complete absence- of cycles in networks having small connectivities. Actually, the region $T=0, F=1, O=1$ is only occupied by $DAG$s. Results are shown in figure (2b-c) of the main text {\bf Hierarchy in complex systems: the possible and the actual}.

Below we proceed to detail the exact parameters of the numerical experiments.

{\bf Directed ER graphs}.- We begin with the nets generated by the directed version of the Erd\"os  R\'enyi model. Such a model consists of two parameters, $|V|$ and $p$, that specify the size of the network ($|V|$) and the probability of connecting any pair by an arc in any direction ($p$). For $|V|\gg 0$, this model does not ensure connectedness until values of $p\approx\log(|V|)/|V|$ \cite{Bollobas:1985}. An ensemble of $35,760$ graphs with three sizes ($|V|=\{100, 250, 500\}$) was created following the following procedure:  We  create $64$ replicas for every graph size starting from  $p_{max}=0.1$. We repeated this process reducing $p$ in a step size of $0.0005$ until a value of $p$  able to produce graphs with a fraction of vertices belonging to the giant connected component larger than $97.5 \% $. The resulting ensemble encompasses a range of  $100<\langle k \rangle < 2.51$. 

{\bf Directed Callaway graphs}.- The uniform attachment \cite{Callaway:2001} is an iterative model consisting of three parameters, namely $m_0$, $m$ and $i$. Starting on a fully connected set of $m_0$ nodes, during $i$ iterations a new node is added to the network. Such node is linked to $m$ existing nodes. These nodes can be selected with equal probability, i.e. there is a {\em uniform} attachment. Once the iterative process is finished, a random uniform orientation that converts links into directed arcs is performed in order to obtain a directed graph. An ensemble of $768$ directed graphs was generated considering $64$ replicas of networks for three different sizes($|V|=\{100, 250, 500\}$) and four averages degrees ($\langle k \rangle= \{2, 4, 6, 8\}$). Average degrees correspond with $m=\{1, 2, 3, 4\}$; in each case, the seed is a clique of $m_0=m$ nodes. 

{\bf Directed BA graphs}.- The preferential attachment \cite{Barabasi:1999} is an iterative model consisting of three parameters, namely $m_0$, $m$ and $i$. Starting on a fully connected set of $m_0$ nodes, during $i$ iterations a new node is added to the network. Such node is linked to $m$ of the existing nodes. These nodes are subject to be selected with a probability proportional to their degree, i.e. there is a preferential attachment.  Once the iterative process is finished, a random uniform orientation that converts links into directed arcs is performed in order to obtain a directed graph. An ensemble of  $768$ directed graphs containing  $64$ replicas of networks for three different sizes($|V|=\{100, 250, 500\}$) and four averages degrees ($\langle k \rangle= \{2, 4, 6, 8\}$, corresponding to $m=\{1, 2, 3, 4\}$) was generated. 
As above, the seed is a clique of $m_0=m$ nodes in all cases.

\subsection{Real Networks}
\label{Sec:RealNetworks}

After looking at the properties of standard models of random networks, we explored a collection of $125$ networks encompassing $13$ types of systems obtained from real data. Figure Results are shown in figure (2d) of the main text {\bf Hierarchy in complex systems: the possible and the actual}, shows the resulting $TFO$ morphospace for the real networks used in this work.  To discuss the relevance of the observed results, we  confronted data of real networks against the one obtained from their randomized counterparts. Randomized ensembles were built using two different techniques of randomization. Below methods and results are detailed, after the network data set presentation.
%

\subsubsection{Network dataset}
\label{Sec:Network dataset}

\subparagraph{\textit{C elegans} cell lineage network.}
Label: Cellular in the Fig 2 of the article.  Cell lineage network obtained from Worm database and pulished as analysed in \cite{Goni:2010}.

\subparagraph{\textit{C elegans} neural network.}
Label: Neuronal in the Fig 2 of the article. A directed, weighted network representing the neural network of {\em C. elegans}. Data compiled by D. Watts and S. Strogatz and made available at the Mark Newman's website: http://www-personal.umich.edu/\~mejn/netdata/ used in \cite{Watts:1998}. Original experimental data taken from \cite{White:1986}.

\subparagraph{Metabolic networks.}
Label: Metabolisms  in the Fig 2 of the article.. Metabolic network set is a selection of 19 reaction-metabolite directed networks obtained from three different published papers for different organisms:

Barabasi dataset for \textit{E.coli}, \textit{B. subtilis}, \textit{S cerevisiae}. http://www.nd.edu/\~networks/resources/metabolic/index.html from data used in \cite{Jeong:2000}.

Networks obtained from the database published in \cite{Ma:2003}. A selection of 5 multicellular animals: \textit{H. sapiens} (hsa), \textit{M. musculus} (mmu), \textit{D. melanogaster} (dme), \textit{R. novergicus} (rno) and \textit{C. elegans} (cel). One multicelular plant: \textit{A. thaliana} (ath). Two unicellular funghi: \textit{S. cerevisiae} (sce), \textit{S. pombe} (spo). And seven prokaryotes: \textit{P. Aeruginosa} (pae), \textit{E. coli} (eco), \textit{B subtilis} (bsu), \textit{Mycoplasma genitalis} (mge), \textit{Mycoplasma pneumoniae} (mpn), \textit{Synechocystis sp.} PCC6803 (syn) and \textit{Salmonella typhimurium} (sty).

The Edinburgh human metabolic network taken from \cite{Ma:2007}.

\subparagraph{Food Webs.}
 Label: Food webs in the Fig 2 of the article. These networks describes the exchange flow from the donor to recipient compartments where not necessarily nodes represent single species.
The food web set contains a set of 22 directed weighted graphs in their original format. In this article were taken as unweighted directed graphs. Food-webs were originally selected from the R.E. Ulanowicz's Collection from the Ecosystem Network Analysis site and from ATLSS - Network Analysis of Trophic Dynamics in South Florida Ecosystems and compiled in the Pajek dataset. Source: http://vlado.fmf.uni-lj.si/pub/networks/data/bio/
foodweb/foodweb.htm.

\subparagraph{Gene regulatory networks (GRNs) and one kinase network.}
 Label: GRNs and Kinase network respectively, in the Fig 2 of the article. GRNs set contains the networks from two different articles. The data consists of directed networks where nodes are genes and arcs between two genes captures the interaction of the respective gene product -represented by the source node- over the regulatory region of the target gene -target node.
The only one available kinase network at the present is represented by  a directed network where nodes are proteins and arcs represents the relation of phosphorylation of a protein -node source- over the a target protein -node target. 
GRNs of \textit{S cerevisiae}, \textit{E. coli}, \textit{Mus musculus}, \textit{Ratus novergicus}, \textit{Homo sapiens}  and \textit{Micoplasma tuberculosis} and one kinase network of \textit{S cerevisiae} were taken  from http://info.gersteinlab.org/Hierarchy and published in \cite{Nitin:2010}.
GRNs of  \textit{S. cerevisiae}, \textit{E. coli}, and \textit{Bacillus subtilis} taken from \cite{Rodriguez-Caso:2009}.

\subparagraph{Electronic circuits.}
Label: Electronic circuits  in the Fig 2 of the article. Directed networks of 50 electronic wiring compiled from ISCAS'89 and ITC'99 sets. Data from electronic circuits used in \cite{Ferrer:2001}.

\subparagraph{Word corpora.}
 Label: Word corpora in the Fig 2 of the article.Four directed lexical graph from different texts according the description in \cite{Ferrer:2001b}: "Angie's Wren Xmas Tale" (Paul Auster), "Frankenstein or the modern prometeous" (Mary Shelley)," Ulisses" (James Joyce), "Moby dick" (Herman Melville), "Black cat" (Edgar Allan Poe) and a fragment of the New York Times column.

\subparagraph{Scientific citation network.}
Label: Citations in the Fig 2 of the article. A directed and feedforward graph of citations networks. Nodes are articles and every article points to the articles in which it is cited. Papers that cite S. Milgram's 1967 Psychology Today paper or use Small World. Taken from pajek dataset: http://vlado.fmf.uni-lj.si/pub/networks/data/

\subparagraph{Software networks or dependence networks.}
Label: Software  in the Fig 2 of the article. Ten directed networks capturing the map of file dependencies in different software. In these networks every node is a file. If a file A is called from another file B, then B receives an arrow from A. Although is generally avoided in programming, one file can content more than one function -argument- and therefore it can be called for more than one purposes. This feature would explain the appearance of cycles in some circumstances.
Eight networks were taken from \cite{Valverde:2005}.
Two additional networks were obtained from the Internet: r package dependencies: taken from http://csgillespie.wordpress.com/2011/03/23/graphical-display-of-r-packages-dependencies/
Java packages taken from: http://gd2006.org/contest/details.php\#java

\subparagraph{Ownership:EVAownership.}
Label: Ownership  in the Fig 2 of the article. EVA is a multidisciplinary research project combining information extraction, information visualization, and social network analysis techniques to bring greater transparency to the public disclosure of inter-relationships between corporations. This data corresponds with an ownership network with 6,726 relationships among 8,343 companies. http://denali.berkeley.edu/eva/
Data obtained form pajek database: http://vlado.fmf.uni-lj.si/pub/networks/data/
econ/Eva/Eva.htm
K. Norlen, G. Lucas, M. Gebbie, and J. Chuang. EVA: Extraction, Visualization and Analysis of the Telecommunications and Media Ownership Network. Proceedings of International Telecommunications Society 14th Biennial Conference (ITS2002), Seoul Korea, August 2002.

\subparagraph{Blogspol.}
Label: Social  in the Fig 2 of the article. A directed network of hyperlinks between weblogs on US politics, recorded in \cite{Adamic:2005}.
Data obtained from pajek database. Original data was taken by pajek database from the Mark Newman's dataset http://www-personal.umich.edu/~mejn/
netdata/ 

\subparagraph{Hitech.}
Label: Social  in the Fig 2 of the article. The case is a small hi-tech computer firm which sells, installs, and maintains computer systems. The network contains the friendship ties among the employees, which were gathered by means of the question: Who do you consider to be a personal friend? A friendship choice (arc) is only included in the network if both persons involved acknowledge it.
Taken from  http://vlado.fmf.uni-lj.si/pub/networks/data/esna/hiTech.htm, published in \cite{Krackhardt:1999}.

\subparagraph{ModMath.}
Label: Social  in the Fig 2 of the article. Taken from Pajek dataset http://vlado.fmf.uni-lj.si/pub
/networks/data/esna/modMath.htm. This network concerns the diffusion of a new mathematic method in the 1950s. This innovation was instigated by top mathematicians and sponsored by the National Science Foundation of the USA as well as the U.S. Department of Education.
R.O. Carlson, Adoption of Educational Innovations (Eugene: University of Oregon, Center for the Advanced Study of Educational Administration, 1965, p. 19).

\subparagraph{PhD advisors (genealogy).}
Label: Genealogies  in the Fig 2 of the article. Data taken from: http://vlado.fmf.uni-lj.si/
pub/networks/data/esna/CSPhD.htm. The network contains the ties between Ph.D. students and their advisors in theoretical computer science; each arc points from a supervisor to a student. The partition contains the (estimated) year in which the Ph.D. was obtained.
Original author: David Johnson; maintained by Ian Parberry. The SIGACT Theoretical Computer Science Genealogy, Last Updated July 22, 1996. Data taken as appeared in \cite{Goni:2010}


%
\subsubsection{Randomization Methods}

\subparagraph{Method A: Preserving the undirected degree sequence and the component structure.}

In this randomization technique, we generate the ensemble of randomized graphs by taking as topological invariants the undirected degree sequence and the component structure. It is worth to note that the component structure is not preserved using a standard method of graph randomization under the so-called {\em configuration model} approach \cite{Newman:2001}.

\begin{figure}[h]
\begin{center}
\includegraphics[width=8cm]{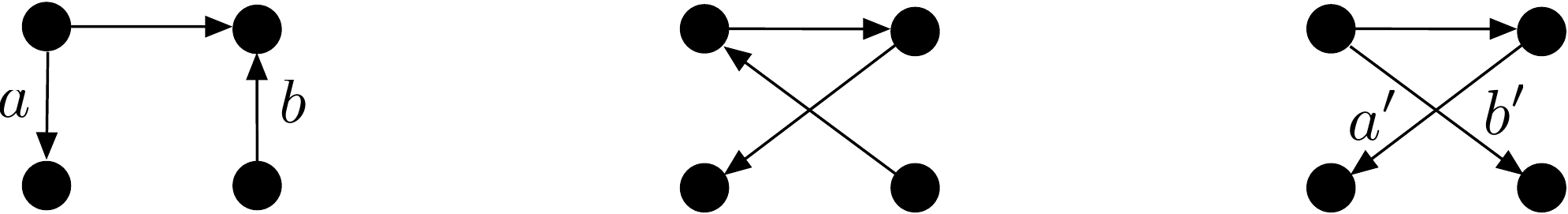}
\caption{A local swap algorithm keeping the undirected degree sequence invariant.}
\label{fig:LocalSwapUnd}
\end{center}
\end{figure}

Given an undirected graph ${\cal G}(V,E)$, $|V|=n$, its {\em undirected degree sequence}, $\sigma_u({\cal G})$, is the sequence of $n$ integer numbers in which the $i$-th number depicts the undirected degree  of node  $v_i$, $k(v_i)$ in a given labeling $v_1,...,v_n$ of the nodes of the graph:
\[
\sigma_u({\cal G})=k(v_1),...,k(v_i),...,k(v_n).
\]
Let us suppose that such a graph has a given component structure $\Gamma({\cal G})={\gamma}_1,...,{\gamma}_j$. The randomization method applied here keeps invariant both $\sigma_u({\cal G})$ and $\Gamma({\cal G})$. This is performed using a {\em directed} version of the {\em local swap} algorithm \cite{Hanhijarvi:2009, Goni:2010}. {\em Local swap} diverges from standard random rewiring algorithms because it keeps invariant the component structure of the graph. The figure below shows how we generate the randomized elements of the ensemble. Let ${\cal G}$ be a directed  graph. We choose three links such that they form a chain of length three in ${\cal G}_u$, figure (\ref{fig:LocalSwapUnd}, left), i.e., a "$\sqsubset$" structure, no matter the direction of the arrows. Then, we cross the links at the extremes of the chain, i.e., generating a 
"$\ltimes$" structure, figure (\ref{fig:LocalSwapUnd} center). Finally, we flip the senses of the arrows at random and, thus, links $a$ and $b$ have transformed into $a'$ and $b'$. If $a'$ or $b'$ previously existed,  we abort this randomizing event and we restart looking at random for another $\sqsubset$ structure within the graph.

For every real network we generated an ensemble of $100$ replicas obtained after applying the randomization algorithm until we performed $4|E|$ link switches or $20|E|$ trials. The latter applies generally if the net belongs to an ensemble having a few members, which can happen due to several reasons, mainly, if the net is too dense or if the net is very sparse, or if it has a non-standard component structure. Numerical details of the randomization process for each real network are described in the Appendix.  

\subparagraph{Method B: Preserving the directed degree sequence and the component structure.}

Now we generate the ensemble of randomized graphs by taking as topological invariants the directed degree sequence and the component structure. 
Given a directed graph ${\cal G}(V,E)$, $|V|=n$, its {\em directed degree sequence}, $\sigma({\cal G})$, is the sequence of $n$ pairs of integer numbers in which the first number of the $i$-th pair depicts the $in$-degree of node and the second one depicts the $out$-degree of node $v_i$, in a given labeling $v_1,...,v_n$ of the nodes of the graph:
\begin{widetext}
\[
\sigma({\cal G})=\langle k_{in}(v_1),k_{out}(v_1)\rangle...,\langle k_{in}(v_i),k_{out}(v_i)\rangle...,\langle k_{in}(v_n), k_{out}(v_n)\rangle.
\]
\end{widetext}
Let us suppose that such a graph has a given component structure $\Gamma({\cal G})={\gamma}_1,...,{\gamma}_j$. The randomization method applied here keeps invariant both $\sigma({\cal G})$ and $\Gamma({\cal G})$. The conservation of $\sigma({\cal G})$ makes this randomization method slightly more restrictive than the other presented above. 
\noindent
Specifically, we look for $\sqsubset$ structures in ${\cal G}$ to perform a {\em Local Swap}, but, now, the sense of the arrows matter. Indeed, the kind of $\sqsubset$  structures over which we can apply a local swap keeping the directed degree sequence of the graph are:
\begin{figure}[h]
\begin{center}
\includegraphics[width=4.2cm]{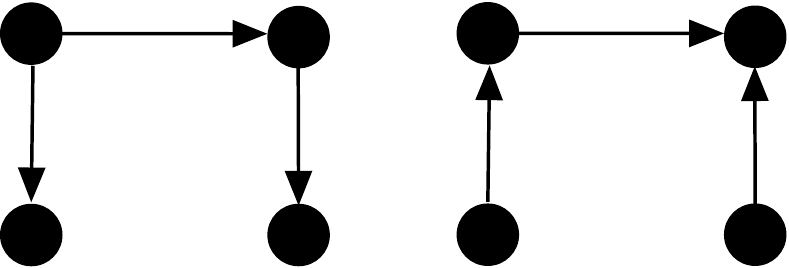}
\caption{Structures over which a  directed  {\em Local Swap} operation can be applied keeping the directed degree sequence of the graph invariant.}
\label{fig:StrucsLocalSwapDir}
\end{center}
\end{figure}

The kind of $\sqsubset$ structures over which we cannot apply  a  directed  {\em Local Swap} operation, if we want to keep the directed degree sequence of the graph invariant are:
\begin{figure}[h]
\begin{center}
\includegraphics[width=4.2cm]{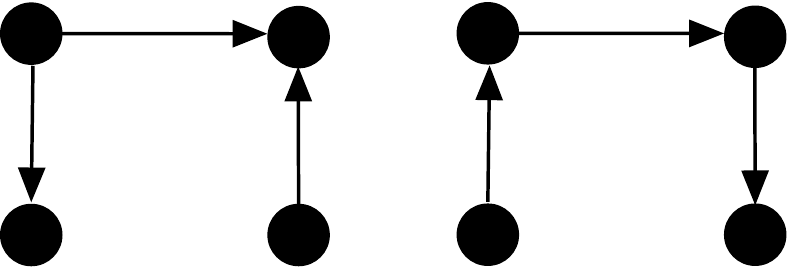}
\caption{Structures over which a  directed  {\em Local Swap} operation cannot be applied in this case.}
\label{fig:StrucsNoLocalSwapDir}
\end{center}
\end{figure}

\begin{figure}[h]
\begin{center}
\includegraphics[width=8cm]{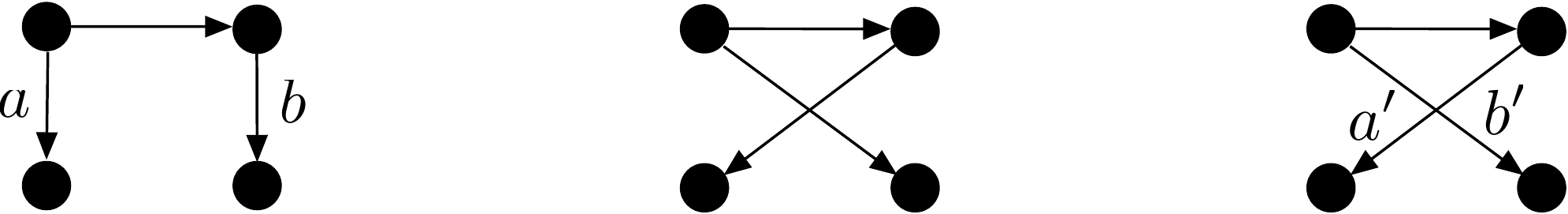}
\caption{A local swap algorithm keeping the directed degree sequence invariant.}
\label{fig:LocalSwapDir}
\end{center}
\end{figure}

Figure (\ref{fig:LocalSwapDir}) shows how we generate the randomized elements of the ensemble. First, we select at random a structure of the kind described in figure (\ref{fig:StrucsLocalSwapDir}), figure  (\ref{fig:LocalSwapDir}, left). Then, we cross the links at the extremes of the chain, i.e., generating a 
"$\ltimes$" structure (\ref{fig:LocalSwapDir} center).  Finally, we check if new links $a'$ and $b'$ previously existed. If so,  we abort this rewiring event and we restart another one looking at random for structures like the ones described in figure (\ref{fig:StrucsLocalSwapDir}).

For every real network we generated an ensemble of $100$ replicas obtained after applying the randomization algorithm until we performed $4|E|$ link switches or $20|E|$ trials, if the net belongs to an ensemble having a few members.
\begin{figure}
\begin{center}								
\includegraphics[width=8cm]{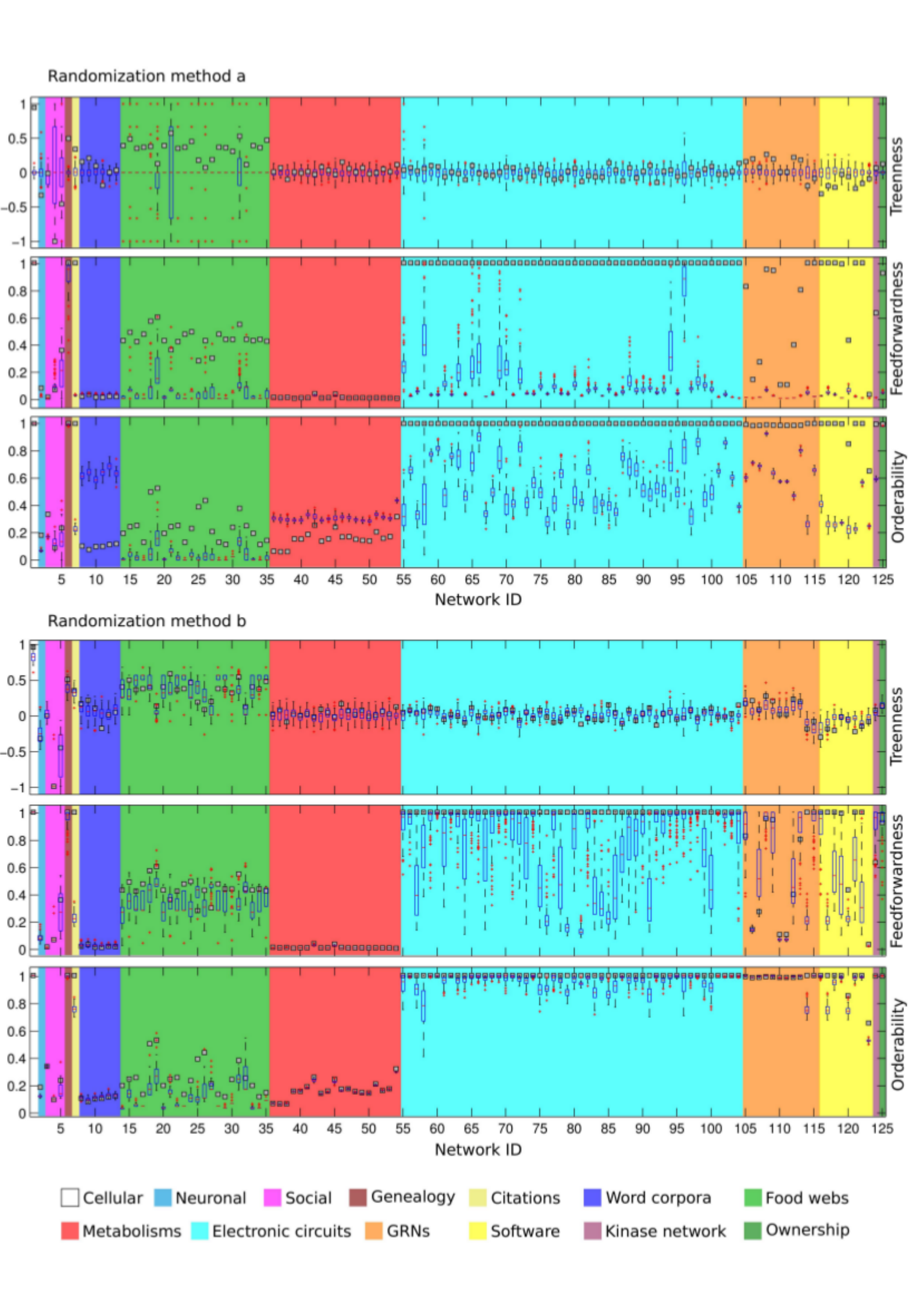}			%
 \caption{Comparison of real networks with randomized ensembles according methods {\bf a} (upper part) and {\bf b} (bottom part). In two groups of three charts, the figure displays three box plots for every randomization method showing the comparison of $T, F, O$ values of the real value with its respective randomized ensemble. X axis labels networks ordered by network ID.} 
 \label{Fig:randomization_composition}	
\end{center}	
\end{figure}

\subsubsection{Confronting real data with their randomized counterparts}

Network randomization offers a quantitative picture of how far real networks are from their respective randomly generated ensembles whose correlations are eliminated by an iterative process of arc rewiring. Comparison is given by contrasting the $TFO$ coordinates, $\mathbf{u}$, of a particular real network with a distribution of values obtained from a randomized ensemble represented by its percentiles in  a box plot fashion. This chart allows us to elude any assumption on the statistical behavior of the ensembles of randomized graphs.  Figure (\ref{Fig:randomization_composition}) shows the comparison between real and random ensemble data, the latter generated according methods {\bf a} and  {\bf b}.  Studied networks  appear clustered by types, and every type of network has an associated color. Numbers, network labels, colors and other additional network quantifiers are detailed in the supplemental information II. In the figure, real network values are represented in charts by a grey square. The distribution of $TFO$ values for every generated ensembles is represented  in a box plot.  Percentiles $25^{\rm th}$, and $75^{\rm th}$ are represented by an empty blue line box  while the percentile $50^{\rm th}$ is depicted with a red line inside. Whiskers show the values within percentiles $10^{\rm th}$ and $90^{\rm th}$. Finally,  red crosses display the values out for the whisker range.

In figure (\ref{Fig:randomization_composition}) looking at the position of  real networks, they fall out from the percentile $50^{\rm th}$ of their randomized ensembles in the $\Omega$ space. Therefore, we can argue that real networks of under scrutiny have not a representative graphical configuration for  their degree sequence. This indicates that, although real networks tend to live in the null model regions as we saw earlier, system's constraints confer differences in the pattern of connections that distinguish real networks from their randomized counterparts. This difference is more accentuated for method {\bf a} (the directed degree sequence is not conserved) than for method {\bf b},  mainly  due to the fact that the latter imposes a more dramatic restriction in the graphical configuration.

In relation to $T$ coordinate, graphs tend to show $T$ values close to zero. However, there are exceptions. Positive biased values are observed in food webs and most of GRNs. Such nets are thus hierarchic ($T>0$), in terms of how the causal flow is organized. On the contrary, electronic circuits show an anti hierarchical configuration ($T<0$), with a tendency to occupy slightly negative, but significant, values of $T$.  Looking at $F$ and $O$ values, real networks are generally far from the whiskers of the ensemble, when confronting real data against the ensemble obtained using the randomization method {\bf a} -where directed degree sequence is preserved. In most cases real data appears to avoid the general cyclic character observed in randomized networks. This trend seems to be reduced when the directed degree sequence is conserved (randomization method {\bf b}). This points to the conclusion that not local correlation among connectivities but the pattern of inputs and outputs would explain part of $TFO$ values. In general, such a trend presents a larger cyclic character in randomizations than in real networks. However, we can observe two exceptions to this general behavior. Looking at the $O$ for randomized ensembles obtained through method {\bf a}, metabolisms and word corpora show a higher cyclic character than their respective randomized ensembles. However, when compared to randomized ensembles obtained using method {\bf b}, the trend is similar to the common values observed in randomized versions. As it happened in the previous cases,  the range occupied by  real and randomized data seems to be fairly justified in most cases by directed degree sequence.


\subsection{The accessibility of the Morphospace: Evolution}
\label{Sec:AccessibilityEvolution}
Now we shall concern ourselves to an {\em in silico} experiment based on an evolutionary algorithm applied to graph population variation. So far we studied how both real nets and their corresponding random ensembles cover the different regions of the morphospace $\Omega$. The numerical experiment presented below will give us important information on how {\em accessible} are the different regions of the morphospace if we impose selective pressures related to specific values of hierarchy coordinates over the possible graphs. In raw words, we let evolve a {\em population} of graphs taking as selective pressure the distance of such graphs to a given target point of $\Omega$. For a given graph, the closer it is to this target point, the higher is its selective value.  At the global, evolutionary  level, what we want to evaluate is how long it takes is to reach a given target point: The more it takes the population to reach it, the more {\em inaccessible} is this region considered. Results are shown in figure (4) of the main text {\bf Hierarchy in complex systems: the possible and the actual}.

Let us detail the experiment. Using a method described in \cite{Marin:1999} we quantified the accessibility of a grid of points in the defined 3D morphospace $\Omega$. This experiment has a high computational cost, thus, we covered $\Omega$  by a grid containing $100$ points in three cuts, corresponding to three $TF$ planes at three different values of orderability. The three planes are the ones defined by $O=\{0.15, 0.5, 0.85\}$. For each decided $O$ value, along the $TF$ plane, minimum and maximum coordinates were distanced a value of $0.05$ distance units\footnote{Given two points  $\mathbf{u}_1=(T_1,F_1,O_1) ,\mathbf{u}_2=(T_2,F_2,O_2)\in \Omega$, the distance is computed using the standard euclidean norm, namely:
\[
d(\mathbf{u}_1,\mathbf{u}_2)=\sqrt{(T_1-T_2)^2+(F_1-F_2)^2+(O_1-O_2)^2}.
\]} 
from the boundaries and all points of the grid were equally distanced at $0.10$. Each of the resulting $TFO$ points was used as a target to be reached or approached by a euclidean distance in the 3D morphospace. We say that to {\em reach} a given target $TFO$  point , $\mathbf{u^*}$, is done when at least one graph of an evolving population acquires a $TFO$ value $\mathbf{u}$ such  that  the euclidean distance between $\mathbf{u}$ and $\mathbf{u^*}$ is  smaller than $0.05$ distance units.
The initial condition of this algorithm is defined by a population of graphs $P$ (always single connected components), being $|P|=25$,  each one made of $|V|=25$ nodes,  generated following the directed Erd\"os  R\'enyi model with $p=0.08$ (see section \ref{Model networks} for model details). 

At each evolutionary iteration  (generation) we perform the following steps:
\begin{itemize}
\item
{Computation of the  $TFO$ values $\mathbf{u}({\cal G})$ for all ${\cal G}\in P$.}
\item
{For each graph ${\cal G}$ of $P$, we compute the euclidean distance $d(\mathbf{u}(({\cal G}),\mathbf{u^*})$ between  $\mathbf{u}({\cal G})$ and desired target point $\mathbf{u^*}\in\Omega$.}
\item
{We calculated the average of the euclidean distance over the population $\langle d(\mathbf{u}({\cal G}) ,\mathbf{u^*})\rangle$.} 
\item
{ We applied the selection criterion: we eliminate from the $P$ those graphs satisfying $d(\mathbf{u}(({\cal G}),\mathbf{u^*})>\langle d(\mathbf{u}({\cal G}),\mathbf{u^*})\rangle$. The set of {\em survivals} is $P'=\{{\cal G}\in P: d(\mathbf{u}({\cal G}),\mathbf{u^*})\leq \langle d(\mathbf{u}({\cal G}),\mathbf{u^*})\rangle\}$. }
\item
{ We create a sequence $\mathbf{s}_{P'}$ picking at random and copying elements from $P'$. We perform this operation  $|P\setminus P'|$ times -therefore, the sequence has length $\ell(\mathbf{s}_{P'})=|P\setminus P'|$, and repetitions are allowed. }
\item
{We apply the {\em randomization operator} over all graphs of the sequence, keeping the graphs of $P'$ invariant. After this step, together with the unchanged graphs excluded from the rewiring process, we obtaining the new generation of $P$.}
\end{itemize} 
This evolutionary algorithm  is then performed a number of $1000$ generations. The {\em randomization} operator is applied to every graph of the sequence $\mathbf{s}_{P'}$ and consists in two steps:
\begin{itemize}
\item
{Random addition from one up to three arcs.}
\item
{Random deletion from one up to three arcs of a graph.  This step must satisfy the condition of preserving the single connected component. If not, randomization event is aborted and new arcs are chosen to be removed.}
\end{itemize}

There is one situation defined by two conditions for which evolutionary algorithm can be stacked:
Given  a generation, $i)$ if none of the graphs become close enough to the target point and $ii)$ none or all graphs are below the fitness mean of the population. Then, to avoid the evolutionary algorithm freezes, the following rule is applied:
\begin{itemize}
\item

Half of the population is chosen at random to be eliminated and replaced by the other half, giving rise to $P'$. 
\end{itemize}

For each target point of the grid, $250$ replicas of the evolutionary experiment were performed. Each population pursuing a target point was allowed to evolve up to $1000$ generations. The average number of generations needed to reach each target point was used as an estimator of the accessibility of that part of $\Omega$. Those cases were the accessibility value is exactly $1000$ indicate that none of the graphs in none of the $250$ processes was able to reach that target point -see figure (4) of the main text for results. In these cases, we expect that more than $1000$ generations are needed.

A note of caution must be added to interpret the strength of the obtained results. It is worth to note that parameters of the evolutionary algorithm may be determinant in shaping the observed fitness landscape, thereby accelerating or slowing the convergence process to the desired target point. The aim of this experiment was just to shed light on how accessible different regions of $\Omega$ are, considering selection and evolution driving forces. In this context, we numerically answered this question choosing a combination of parameters which we consider to be enough representative to provide us relevant information. And indeed, thanks to this approximation, we showed that not all regions are equally accessible but, instead, some regions are fairly improbable to be achieved. Among other factors, the size of the network may play a crucial role to achieve certain regions of the morphospace, since, for example it is easier for small networks to display extreme configurations.  


The described results create a nice, global picture of what is possible and what is actually observed within the zoo of complex networks. This objective is achieved from a rigorous definition of hierarchy.


\end{document}